\documentclass[12pt,preprint]{aastex}
\usepackage{graphics, ulem}
\input psfig.sty
\input epsf
\begin{document}
\baselineskip=12pt

\title{INTERSTELLAR TURBULENCE I: OBSERVATIONS AND PROCESSES}
\markboth{\sc Elmegreen \& Scalo}{\sc Interstellar Turbulence I}

\author{Bruce G. Elmegreen \affil{IBM Research Division, Yorktown Heights, New York 10598; email:
bge@watson.ibm.com}}

\author{John Scalo \affil{Department of Astronomy, University of Texas,
Austin, Texas 78712; e-mail: parrot@astro.as.utexas.edu}}


\begin{abstract}

Turbulence affects the structure and motions of nearly all
temperature and density regimes in the interstellar gas. This
two-part review summarizes the observations, theory, and
simulations of interstellar turbulence and their implications for
many fields of astrophysics. The first part begins with
diagnostics for turbulence that have been applied to the cool
interstellar medium, and highlights their main results. The energy
sources for interstellar turbulence are then summarized along with
numerical estimates for their power input. Supernovae and
superbubbles dominate the total power, but many other sources
spanning a large range of scales, from swing amplified
gravitational instabilities to cosmic ray streaming, all
contribute in some way. Turbulence theory is considered in detail,
including the basic fluid equations, solenoidal and compressible
modes, global inviscid quadratic invariants, scaling arguments for
the power spectrum, phenomenological models for the scaling of
higher order structure functions, the direction and locality of
energy transfer and cascade, velocity probability distributions,
and turbulent pressure.  We emphasize expected differences between
incompressible and compressible turbulence. Theories of magnetic
turbulence on scales smaller than the collision mean free path are
included, as are theories of magnetohydrodynamic turbulence and
their various proposals for power spectra. Numerical simulations
of interstellar turbulence are reviewed. Models have reproduced
the basic features of the observed scaling relations, predicted
fast decay rates for supersonic MHD turbulence, and derived
probability distribution functions for density. Thermal
instabilities and thermal phases have a new interpretation in a
supersonically turbulent medium. Large-scale models with
various combinations of
self-gravity, magnetic fields, supernovae, and
star formation are beginning to resemble the observed interstellar medium
in morphology and statistical properties. The role of self-gravity
in turbulent gas evolution is clarified, leading to new paradigms
for the formation of star clusters, the stellar mass function, the
origin of stellar rotation and binary stars, and the effects of
magnetic fields. The review ends with a reflection on the progress
that has been made in our understanding of the interstellar
medium, and offers a list of outstanding problems.

\end{abstract}

\section{INTRODUCTION}
\label{sect:intro}

In 1951, von Weizs\"acker (1951) outlined a theory for
interstellar matter (ISM) that is similar to what we believe
today: cloudy objects with a hierarchy of structures form in
interacting shock waves by supersonic turbulence that is stirred
on the largest scale by differential galactic rotation and
dissipated on small scales by atomic viscosity. The ``clouds''
disperse quickly because of turbulent motions, and on the largest
scales they produce the flocculent spiral structures observed in
galaxies. In the same year, von Hoerner (1951) noticed that rms
differences in emission-line velocities of the Orion nebula
increased with projected separation as a power law with a power
$\alpha$ between 0.25 and 0.5, leading him to suggest that the gas
was turbulent with a Kolmogorov energy cascade (for which $\alpha$
would be $0.33$; Section 4.6). Wilson et al. (1959) later got a
steeper function, $\alpha\sim0.66$, using better data, and
proposed it resulted from compressible turbulence. Correlated
motions with a Kolmogorov structure function (Section
\ref{sect:diag}) in optical absorption lines were observed by
Kaplan (1958). One of the first statistical models of a continuous
and correlated gas distribution was by Chandrasekhar \& M\"unch
(1952), who applied it to extinction fluctuations in the Milky Way
surface brightness. Minkowski (1955) called the ISM ``an entirely
chaotic mass ... of all possible shapes and sizes ... broken up
into numerous irregular details.''

These early proposals regarding pervasive turbulence failed to
catch on. Interstellar absorption and emission lines looked too
smooth to come from an irregular network of structures -- a
problem that is still with us today (Section 2). The extinction
globules studied by Bok \& Reilly (1947) looked too uniform and
round, suggesting force equilibrium. Oort \& Spitzer (1955) did
not believe von Weizs\"acker's model because they thought galactic
rotational energy could not cascade down to the scale of cloud
linewidths without severe dissipation in individual cloud
collisions. Similar concerns about dissipation continue to be
discussed (Sections \ref{sect:heating}, \ref{sect:decay}). Oort
and Spitzer also noted that the ISM morphology appeared wrong for
turbulence: ``instead of more or less continuous vortices, we find
concentrated clouds that are often separated by much larger spaces
of negligible density.'' They expected turbulence to resemble the
model of the time, with space-filling vortices in an
incompressible fluid, rather than today's model with most of the
mass compressed to a small fraction of the volume in shocks
fronts. When a reddening survey by Scheffler (1967) used structure
functions to infer power-law correlated structures up to 5$^\circ$
in the sky, the data were characterized by saying only that there
were two basic cloud types, large (70 pc) and small (3 pc), the
same categories popularized by Spitzer (1968) in his textbook.

Most of the interesting physical processes that could be studied
theoretically at the time, such as the expansion of ionized
nebulae and supernovae (SNe) and the collapse of gas into stars,
could be modeled well enough with a uniform isothermal medium.
Away from these sources, the ISM was viewed as mostly static, with
discrete clouds moving ballistically. The discovery of broad
emission lines and narrow absorption lines in {\it H I} at 21 cm
reinforced this picture by suggesting a warm intercloud medium in
thermal pressure balance with the cool clouds (Clark 1965). ISM
models with approximate force equilibrium allowed an ease of
calculation and conceptualization that was not present with
turbulence. Supernovae were supposed to account for the energy,
but mostly by heating and ionizing the diffuse phases (McCray \&
Snow 1979). Even after the discoveries of the hot intercloud
(Bunner et al. 1971, Jenkins \& Meloy 1974) and cold molecular
media (Wilson et al. 1970), the observation of a continuous
distribution of neutral hydrogen temperature (Dickey, Salpeter, \&
Terzian 1977), and the attribution of gas motions to supernovae
(e.g., McKee \& Ostriker 1977), there was no compelling reason to
dismiss the basic cloud-intercloud model in favor of widespread
turbulence. Instead, the list of ISM equilibrium ``phases'' was
simply enlarged. Supersonic linewidths, long known from {\it H I}
(e.g., McGee, Milton, \& Wolfe 1966, Heiles 1970) and optical
(e.g., Hobbs 1974) studies and also discovered in molecular
regions at this time (see Zuckerman \& Palmer 1974), were thought
to represent magnetic waves in a uniform cloud (Arons \& Max
1975), even though turbulence was discussed as another possibility
in spite of problems with the rapid decay rate (Goldreich \& Kwan
1974, Zuckerman \& Evans 1974). A lone study by Baker (1973) found
large-scale correlations in {\it H I} emission and presented them
the context of ISM turbulence, deriving the number of ``turbulent
cells in the line of sight,'' instead of the number of ``clouds.''
Mebold, Hachenberg \& Laury-Micoulaut (1974) followed this with
another statistical analysis of the {\it H I} emission. However,
there was no theoretical context in which the Baker and Mebold et
al. papers could flourish given the pervasive models about
discrete clouds and two or three-phase equilibrium.

The presence of turbulence was more widely accepted for very small
scales. Observations of interstellar scintillation at radio
wavelengths implied there were correlated structures (Rickett
1970), possibly related to turbulence (Little \& Matheson 1973),
in the ionized gas at scales down to $10^9$ cm or lower (Salpeter
1969; {\it Interstellar Turbulence II,} next chapter, this
volume). This is the same scale at which cosmic rays ({\it
Interstellar Turbulence II}) were supposed to excite magnetic
turbulence by streaming instabilities (Wentzel 1968a). However,
there was (and still is) little understanding of the physical
connection between these small-scale fluctuations and the
larger-scale motions in the cool neutral gas.

Dense structures on resolvable scales began to look more like
turbulence after Larson (1981) found power-law correlations
between molecular cloud sizes and linewidths that were reminiscent
of the Kolmogorov scaling law. Larson's work was soon followed by
more homogeneous observations that showed similar correlations
(Myers 1983, Dame et al. 1986, Solomon et al. 1987). These motions
were believed to be turbulent because of their power-law nature,
despite continued concern with decay times, but there was little
recognition that turbulence on larger scales could also form the
same structures in which the linewidths were measured. Several
reviews during this time reflect the pending transition (Dickey
1985, Dickman 1985, Scalo 1987, Dickey \& Lockman 1990).

Perhaps the most widespread change in perception came when the
Infrared Astronomical Satellite (IRAS) observed interstellar
``cirrus'' and other clouds in emission at 100 $\mu$ (Low et al.
1984). The cirrus clouds are mostly transparent at optical
wavelengths, so they should be in the diffuse cloud category, but
they were seen to be filamentary and criss-crossed, with little
resemblance to ``standard'' clouds. Equally complex structures
were present even in IRAS maps of ``dark clouds,'' like Taurus,
and they were observed in maps of molecular clouds, such as the
Orion region (Bally et al. 1987). The wide field of view and good
dynamic range of these new surveys finally allowed the diffuse and
molecular clouds to reveal their full structural complexity, just
as the optical nebulae and dark clouds did two decades earlier.
Contributing to this change in perception was the surprising
discovery by Crovisier \& Dickey (1983) of a power spectrum for
widespread {\it H I} emission that was comparable to the
Kolmogorov power spectrum for velocity in incompressible
turbulence. CO velocities were found to be correlated over a range
of scales, too (Scalo 1984, Stenholm 1984). By the late 1980s,
compression from interstellar turbulence was considered to be one
of the main cloud-formation mechanisms (see review in Elmegreen
1991).

Here we summarize observations and theory of interstellar
turbulence. This first review discusses the dense cool phases of
the ISM, energy sources, turbulence theory, and simulations. {\it
Interstellar Turbulence II} considers the effects of turbulence on
element mixing, chemistry, cosmic ray scattering, and radio
scintillation.

There are many reviews and textbooks on turbulence. A
comprehensive review of magnetohydrodynamical (MHD) turbulence is
in the recent book by Biskamp (2003), and a review of laboratory
turbulence is in Sreenivasan \& Antonia (1997). A review of
incompressible MHD turbulence is in Chandran (2003). For the ISM,
a collection of papers covering a broad range of topics is in the
book edited by Franco \& Carraminana (1999). Recent reviews of ISM
turbulence simulations are in V\'azquez-Semadeni et al. (2000),
Mac Low (2003), and Mac Low \& Klessen (2004), and a review of
observations is in Falgarone, Hily-Blant \& Levrier (2003). A
review of theory related to the ISM is given by V\'azquez-Semadeni
(1999).  Earlier work is surveyed by Scalo (1987). General
discussions of incompressible turbulence can be found in Tennekes
\& Lumley (1972), Hinze (1975), Lesieur (1990), McComb (1990),
Frisch (1995), Mathieu \& Scott (2000), Pope (2000), and Tsinober
(2001). The comprehensive two volumes by Monin \& Yaglom (1975)
remain extremely useful. Work on compressibility effects in
turbulence at fairly low Mach numbers is reviewed by Lele (1994).
Generally the literature is so large that we can reference only a
few specific results on each topic; the reader should consult the
most recent papers for citations of earlier work.

We have included papers that were available to us as of Dec. 2003.
A complete bibliography including paper titles is available at:

http://www.as.utexas.edu/astronomy/people/scalo/research/ARAA/

\section{DIAGNOSTICS OF TURBULENCE IN THE DENSE INTERSTELLAR MEDIUM}
\label{sect:diag}

\begin{figure}
\plotone{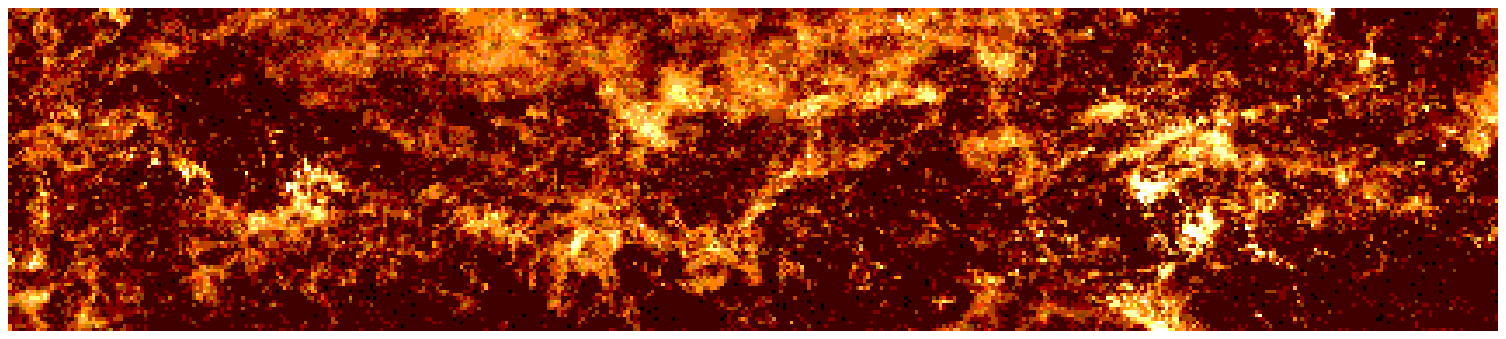} 
\caption{$^{12}$CO emission from
the outer Galaxy spanning a range of Galactic longitude from
102.5$^\circ$ to 141.5$^\circ$ and latitude from -3$^\circ$ to
5.4$^\circ$, sampled every 50'' with a 45'' beam. The emission is
integrated between -110 km s$^{-1}$ and 20 km s$^{-1}$. The high
latitude emission is mostly from local gas at low velocity,
consisting of small structures that tend to be
nonself-gravitating. The low latitude emission is a combination of
local gas plus gas in the Perseus arm at higher velocity. The
Perseus arm gas includes the giant self-gravitating clouds that
formed the OB associations that excite W3, W4, and W5 ({\it left})
and NGC 7538 ({\it right}). Most structures are hierarchical and
self-similar on different scales, so the distant massive clouds
look about the same on this map as the local low-mass clouds. This
image was kindly provided by M. Heyer using data from Heyer et al.
(1998). {\it (for slightly higher resolution, see paperI-f1.gif)}}\label{fig:heyer}
\end{figure}

\begin{figure}
\plotone{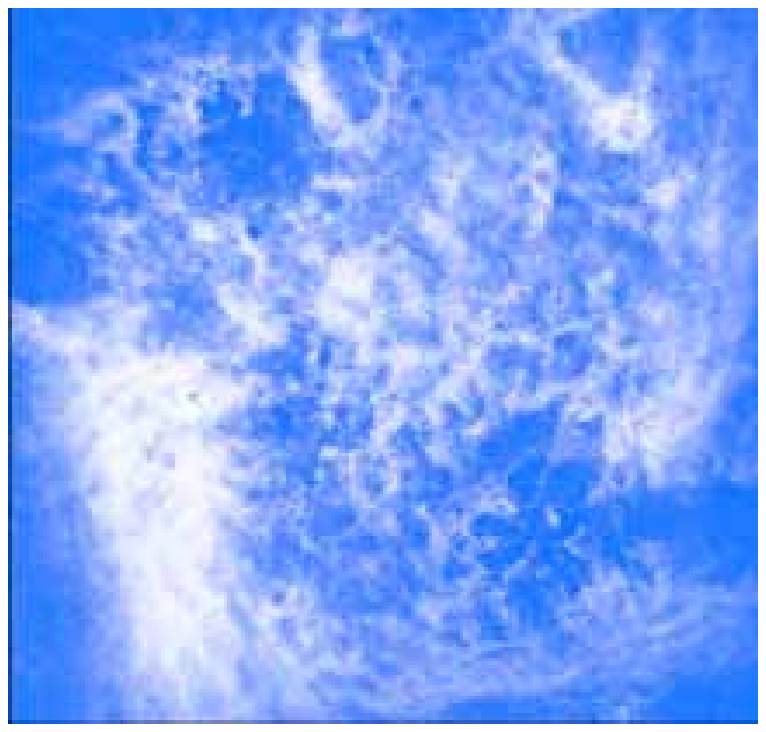} 
\caption{Integrated {\it H I}
emission from the LMC showing shells and filaments covering a wide
range of scales. The 2D power spectrum of this emission is a power
law from 20 pc to 1 kpc with a slope of $\sim-3.2$. Data from
Elmegreen, Kim \& Staveley-Smith
(2001){\it (for slightly higher resolution, see 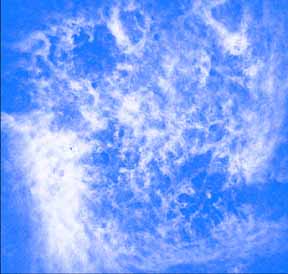)}.\label{fig:lmc}}\end{figure}

The interstellar medium presents a bewildering variety of
intricate structures and complex motions that cannot be compressed
entirely into a few numbers or functions (Figures \ref{fig:heyer}
and \ref{fig:lmc}). Any identification with physical processes or
simulations must involve a large set of diagnostic tools. Here we
begin with analysis techniques involving correlation functions,
structure functions, and other statistical descriptors of column
density and brightness temperature, and then we discuss techniques
that include velocities and emission-line profiles.

Interstellar turbulence has been characterized by structure
functions, autocorrelations, power spectra, energy spectra, and
delta variance, all of which are based on the same basic
operation. The structure function of order $p$ for an observable
$A$ is
\begin{equation}S_p(\delta r)=<|A(r)-A(r+\delta r)|^p>,\label{eq:sf}\end{equation}
for position $r$ and increment $\delta r$, and the power-law fit
to this, $S_p(\delta r)\propto \delta r^{\zeta_p},$ gives the
slope $\zeta_p$. The autocorrelation of $A$ is
\begin{equation}
C(\delta r)=<A(r) A(r+\delta r)>,\label{eq:correl}\end{equation}
and the power spectrum is
\begin{equation}
P(k)= {\hat A(k)}{\hat A(k)}^*
\end{equation}
for Fourier transform ${\hat A}=\int e^{ikr}A(r)dr$ and complex
conjugate $A^*$. The Delta variance (St\"utzki et al. 1998,
Zielinsky \& St\"utzki 1999) is a way to measure power on various
scales using an unsharp mask:
\begin{equation}
\sigma_\Delta^2(L)=< \int_0^{3L/2}
dx\left\{\left(A\left[r+x\right]-<A> \right) \bigodot
\left(x\right)\right\}^2 > \label{eq:dv}
\end{equation}
for a two-step function
\begin{equation}
\bigodot(x)=\pi \left(L/2\right)^{-2}\times\left\{1 \;\; {\rm
for}\;\;x<L/2\;\;,\;\; {\rm and} -0.125 \;\;{\rm for}\;
L/2<x<3L/2\right\}.
\end{equation}
In these equations, the average over the map, indicated by $<>$,
is used as an estimate of the ensemble average.

The power spectrum is the Fourier transform of the autocorrelation
function, and for a statistically homogeneous and isotropic field,
the structure function-of-order $p=2$ is the mean-squared $A$
minus twice the autocorrelation: $S_2=<A^2>-2C$. The delta
variance is related to the power spectrum: For an emission
distribution with a power spectrum $\propto k^{-n}$ for wave number
$k$, the delta variance is $\propto r^{n-2}$ for $r=1/k$
(Bensch, St\"utzki \& Ossenkopf 2001; Ossenkopf et al. 2001).

We use the convention where the energy spectrum $E(k)$ is
one-dimensional (1D) and equals the average over all directions of
the power spectrum, $E(k)dk=P({\bf k})dk^D$ for number of
dimensions $D$ (Section \ref{sect:scale}). For incompressible
turbulence, the Kolmogorov power spectrum in three-dimensions (3D)
is $\propto k^{-11/3}$ and the energy spectrum is $E(k)\propto
k^{-5/3}$; for a two-dimensional (2D) distribution of this
fluctuating field, $P\propto k^{-8/3}$ and in 1D, $P\propto
k^{-5/3}$ for the same $E(k)$. The term energy refers to any
squared quantity, not necessarily velocity.

Power spectra of Milky Way {\it H I} emission (Green 1993, Dickey
et al. 2001, Miville-Desch${\hat {\rm e}}$nes et al. 2003), {\it H
I} absorption (Deshpande, Dwarakanath \& Goss 2000), CO emission
(St\"utzki et al. 1998, Plume et al. 2000, Bensch et al. 2001),
and IRAS 100 $\mu$ emission (Gautier et al. 1992; Schlegel,
Finkbeiner \& Davis 1998) have power-law slopes of around $-2.8$
to $-3.2$ in 2D maps. The same power laws were found for 2D {\it H
I} emission from the entire Small and Large Magellanic Clouds
(SMC, LMC) (Stanimirovic et al. 1999; Elmegreen, Kim \&
Staveley-Smith 2001) and for dust emission from the SMC
(Stanimirovic et al. 2000). These intensity power spectra are
comparable to but steeper than the 2D (projected) power spectra of
velocity in a turbulent incompressible medium, $-8/3$, although
the connection between density and velocity spectra is not well
understood (see discussion in Klessen 2000).

One-dimensional power spectra of azimuthal profiles in galaxies
have the same type of power law, shallower by one because of the
reduced dimension.  This is shown by {\it H I} emission from the
LMC (Elmegreen et al. 2001), star formation spirals in flocculent
galaxies (Elmegreen et al. 2003), and dust spirals in galactic
nuclei (Elmegreen et al. 2002).  A transition in slope from
$\sim-5/3$ on large scales to $\sim-8/3$ on small scales in the
azimuthal profiles of {\it H I} emission from the LMC was shown to
be consistent with a transition from 2D to 3D geometry, giving the
line-of-sight thickness of the {\it H I} layer (Elmegreen et al.
2001).

Power spectra of optical starlight polarization over the whole sky
have power-law structure too, with a slope of $-1.5$ for angles
greater than $\sim10$ arcmin (Fosalba et al. 2002).  A 3D model of
field line irregularities with a Kolmogorov spectrum and random
sources reproduces this result (Cho \& Lazarian 2002a).

Models of the
delta-variance for isothermal MHD turbulence
were compared with observations of the Polaris Flare by
Ossenkopf \& Mac Low (2002). The models showed a flattening of the
delta-variance above the driving scale and a steepening below the
dissipation scale, leading Ossenkopf \& Mac Low to conclude that
turbulence is driven from the outside and
probably dissipated below the resolution limit. Ossenkopf et al.
(2001) compared delta-variance observations to models with and
without gravity, finding that gravitating models produce
relatively more power on small scales, in agreement with 3-mm
continuum maps of Serpens. Zielinsky \& St\"utzski (1999) examined
the relation between wavelet transforms and the delta-variance,
finding that the latter gives the variance of the wavelet
coefficients. The delta-variance avoids problems with map boundaries,
unlike power spectra
(Bench et al. 2001), but it can be dominated by noise
when applied to velocity centroid maps
(Ossenkopf \& Mac Low 2002).

Padoan, Cambr\'esy \& Langer (2002) obtained a structure function
for extinction in the Taurus region and found that
$\zeta_p/\zeta_3$ varies for $p=1$ to 20 in the same way as the
velocity structure function in a model of supersonic turbulence
proposed by Boldyrev (2002). Padoan et al. (2003a) got a similar
result using $^{13}$CO emission from Perseus and Taurus. In
Boldyrev's model, dissipation of supersonic turbulence is assumed
to occur in sheets, giving $\zeta_p/\zeta_3=p/9+1-3^{-p/3}$ for
velocity (see Sections \ref{sect:interm} and \ref{sect:goldreich}).

Other spatial information was derived from wavelet transforms,
fractal dimensions, and multifractal analysis. Langer et al.
(1993) studied hierarchical clump structure in the dark cloud B5
using unsharp masks, counting emission features as a function of
size and mass for filter scales that spanned a factor of eight.
Analogous structure was seen in galactic star-forming regions
(Elmegreen \& Elmegreen 2001), nuclear dust spirals (Elmegreen et
al. 2002), and LMC {\it H I} emission (Elmegreen et al. 2001).
Wavelet transforms were used on optical extinction data to provide
high-resolution panoramic images of the intricate structures
(Cambr\'esy 1999).

Perimeter-area scaling gives the fractal dimension of a contour
map. Values of 1.2 to 1.5 were measured for extinction (Beech
1987, Hetem \& Lepine 1993), {\it H I} emission in high velocity
clouds (Wakker 1990), 100 micron dust intensity or column density
(Bazell \& Desert 1988; Dickman, Horvath \& Margulis 1990; Scalo
1990; Vogelaar \& Wakker 1994), CO emission (Falgarone, Phillips
\& Walker 1991), and {\it H I} emission from M81 group of galaxies
(Westpfahl et al. 1999) and the LMC (Kim et al. 2003). This
fractal dimension is similar to that for terrestrial clouds and
rain areas and for slices of laboratory turbulence.  If the
perimeter-area dimension of a projected 3D structure is the same
as the perimeter-area dimension of a slice, then the ISM value of
$\sim1.4$ for projected contours is consistent with analogous
measures in laboratory turbulence (Sreenivasan 1991).

Chappell \& Scalo (2001) determined the multifractal spectrum,
$f(\alpha)$, for column density maps of several regions
constructed from IRAS 60 $\mu$ and 100 $\mu$ images. The parameter
$\alpha$ is the slope of the increase of integrated intensity with
scale, $F(L)\propto L^\alpha$. The fractal dimension $f$ of the
column density surface as a function of $\alpha$ varies as the
structure changes from point-like ($f\sim2$) to filamentary
($f\sim1$) to smooth ($f=0$).  Multifractal regions are
hierarchical, forming by multiplicative spatial processes and
having a dominant geometry for substructures. The region-to-region
diversity found for $f(\alpha)$ contrasts with the uniform
multifractal spectra in the energy dissipation fields and passive
scalar fields of incompressible turbulence, and also with the
uniformity of the perimeter-area dimension, giving an indication
that compressible ISM turbulence differs qualitatively from
incompressible turbulence.

Hierarchical structure was investigated in Taurus by Houlahan \&
Scalo (1992) using a structure tree. They found linear
combinations of tree statistics that could distinguish between
nested and nonnested structures in projection, and they also
estimated tree parameters like the average number of clumps per
parent.  A tabulation of three levels of hierarchical structure in
dark globular filaments was made by Schneider \& Elmegreen (1979).

Correlation techniques have also included velocity information
(see review in Lazarian 1999). The earliest studies used the
velocity for $A$ in equations \ref{eq:sf} to \ref{eq:dv}. Stenholm
(1984) measured power spectra for CO intensity, peak velocity, and
linewidth in B5, finding slopes of $-1.7\pm0.3$ over a factor of
$\sim10$ in scale.  Scalo (1984) looked at $S_2\left(\delta
r\right)$ and $C\left(\delta r\right)$ for the velocity centroids
of $^{18}$CO emission in $\rho$ Oph and found a weak correlation.
He suggested that either the correlations are partially masked by
errors in the velocity centroid or they occur on scales smaller
than the beam.  In the former case the correlation scale was about
0.3--0.4 pc, roughly a quarter the size of the mapped region.
Kleiner \& Dickman (1985) did not see a
correlation for velocity centroids of $^{13}$CO emission in
Taurus, but later used higher resolution data for Heiles Cloud 2
in Taurus and reported a correlation on scales less than 0.1 pc
(Kleiner \& Dickman 1987).  Overall, the attempts to construct the
correlation function or related functions for local cloud
complexes have not yielded a consistent picture.

P\'erault et al. (1986) determined $^{13}$CO autocorrelations for
two clouds at different distances, noted their similarities, and
suggested that the resolved structure in the nearby cloud was
present but unresolved in the distant cloud. They also obtained a
velocity-size relation with a power-law slope of $\sim0.5$. Hobson
(1992) used clump-finding algorithms and various correlation
techniques for HCO$^+$ and HCN in M17SW; he found correlations
only on small scales ($<1$ pc) and got a power spectrum slope for
velocity centroid fluctuations that was slightly shallower than
the Kolmogorov slope. Kitamura et al. (1993) considered clump
algorithms and correlation functions for Taurus and found no power
law but a concentration of energy on 0.03-pc scales; they noted
severe edge effects, however.  Miesch \& Bally (1994) analyzed
centroid velocities in several molecular clouds, cautioned about
sporadic effects near the beam scale, and found a correlation
length that increased with the map size. They concluded, as in
P\'erault et al., that the ISM was self-similar over a wide range
of scales.  Miesch \& Bally also used a structure function to
determine a slope of $0.43\pm0.15$ for the velocity-size relation.
Gill \& Henriksen (1990) introduced wavelet transforms for the
analysis of $^{13}$CO centroid velocities in L1551 and measured a
steep velocity-size slope, 0.7.

Correlation studies like these give the second order moment of the
two-point probability distribution functions (pdfs). One-point
pdfs give no spatial information but contain all orders of
moments. Miesch \& Scalo (1995) and Miesch, Scalo \& Bally (1999)
found that pdfs for centroid velocities in molecular clouds are
often non-Gaussian with exponential or power law tails and
suggested the physical processes involved differ from
incompressible turbulence, which has nearly Gaussian centroid pdfs
(but see Section \ref{sect:velo}). Centroid-velocity pdfs with fat
tails have been found many times since the 1950s using optical
interstellar lines, {\it H I} emission, and {\it H I} absorption
(see Miesch et al. 1999). Miesch et al. (1999) also plotted the
spatial distributions of the pixel-to-pixel {differences} in the
centroid velocities for several molecular clouds and found complex
structures. The velocity difference pdfs had enhanced tails on
small scales, which is characteristic of intermittency (Section
\ref{sect:interm}). The velocity difference pdf in the Ophiuchus
cloud has the same enhanced tail, but a map of this difference
contains filaments reminiscent of vortices (Lis et al. 1996,
1998). Velocity centroid distributions observed in atomic and
molecular clouds were compared with hydrodynamic and MHD
simulations by Padoan et al. (1999), Klessen (2000), and Ossenkopf
\& Mac Low (2002). Lazarian \& Esquivel (2003) considered a
modified velocity centroid, designed to give statistical
properties for both the supersonic and subsonic regimes and the
power spectrum of solenoidal motions in the subsonic regime.

The most recent techniques for
studying structure use all of the spectral line data, rather
than the centroids alone. These techniques include the spectral
correlation function, principal component analysis, and velocity
channel analysis.

The spectral correlation function $S(x,y)$ (Rosolowsky et al.
1999) is the average over all neighboring spectra of the
normalized rms difference between brightness temperatures. A
histogram of $S$ reveals the autocorrelation properties of a
cloud: If $S$ is close to unity the spectra do not vary much.
Rosolowsky et al. found that simulations of star-forming regions
need self-gravity and magnetic fields to account for the
large-scale integrity of the cloud. Ballesteros-Paredes,
V\'azquez-Semadeni \& Goodman (2002) found that self-gravitating
MHD simulations of the atomic ISM need realistic energy sources,
while Padoan, Goodman \& Juvela (2003) got the best fit to
molecular clouds when the turbulent speed exceeded the Alfv\'en
speed. Padoan et al. (2001c) measured the line-of-sight thickness
of the LMC using the transition length where the slope of the
spectral correlation function versus separation goes from steep on
small scales to shallow on large scales.

Principle component analysis (Heyer \& Schloerb 1997) cross
correlates all pairs of velocity channels, $\left(v_i,v_j\right),$
by multiplying and summing the brightness temperatures at
corresponding positions:
\begin{equation}
S_{i,j}\equiv S\left(v_i,v_j\right)={1\over {n}}\sum_{a=1}^{n}
T([x,y]_a,v_i)T([x,y]_a,v_j);\end{equation} $n$ is the number of
positions in the map.  The matrix $S_{i,j}$ contains information
about the distribution of all emitting velocities, but averages
out spatial information. The analysis uses $S_{i,j}$ to find an
orthogonal normalized basis set of eigenvectors that describes the
velocity distribution. A typical cloud may need only a few
eigenvectors; a field of CO emission in NGC 7538 needed only seven
components before the level of variation was comparable to the
noise (Brunt \& Heyer 2002). Brunt \& Heyer (2002) used this
technique to determine the average scaling between correlations in
velocity and position, obtaining a power law with a slope of
$\sim0.6$ for several CO sources. Their fractal Brownian motion
models of clouds then suggested the intrinsic slope was $\sim0.5$.
More recent studies comparing observations to simulations suggest
the slope is 0.5 to 0.8, which implies a steep energy spectrum,
$E(k)\propto k^{-2}$ to $k^{-2.6}$, although there are some
ambiguities in the method (Brunt et al. 2003).

Velocity channel analysis was developed by Lazarian \& Pogosyan
(2000) to make use of the fact that the power spectrum of emission
from a turbulent gas has a shallower slope in narrow velocity
channels than in wide channels. This is a general property of
correlated velocity fields. The shallow slope is the result of an
excess of small features from unrelated physical structures that
blend by velocity crowding on the line of sight. This blending
effect has been studied and substantially confirmed in a number of
observations (Dickey et al. 2001, Stanimirovic \& Lazarian 2001)
and simulations (Ballesteros-Paredes, V\'azquez-Semadeni \& Scalo
1999; Pichardo et al.  2000; Lazarian et al. 2001; Esquivel et al.
2003).  Miville-Desch${\hat {\rm e}}$nes, Levrier \& Falgarone
(2003) noted that the method would not give the correct intrinsic
power spectrum if the velocity channels were not narrow enough. If
there are no velocity fluctuations, then the power spectrum of
projected emission from a slab that is thinner than the inverse
wave number is shallower than the 3D density spectrum by one
(Goldman 2000, Lazarian et al. 2001).

Emission-line profiles also contain information about interstellar
turbulence. Molecular cloud observations suggest that profile
width varies with size of the region to a power between 0.3 and
0.6 (e.g., Falgarone, Puget \& Perault 1992; Jijina, Myers \&
Adams 1999). Similar scaling was found for {\it H I} clouds by
Heithausen (1996).  However other surveys using a variety of
tracers and scales yield little correlation (e.g., Plume et al.
1997, Kawamura et al. 1998, Peng et al. 1998, Brand et al. 2001,
Simon et al. 2001, Tachihara et al. 2002), or yield correlations
dominated by scatter (Heyer et al. 2001). An example of extreme
departure from this scaling is the high resolution CO observation
of clumps on the outskirts of Heiles Cloud 2 in Taurus, with sizes
of $\sim0.1$ pc and unusually large linewidths of $\sim2$ km
s$^{-1}$ (Sakamoto \& Sunada 2003). Overall, no definitive
characterization of the linewidth-size relation has emerged.

Most clouds with massive star formation have non-Gaussian and
irregular line profiles, as do some quiescent clouds (MBM12---
Park et al. 1996, Ursa Majoris---Falgarone et al. 1994, Figure
2b,c), while most quiescent clouds and even clouds with moderate
low-mass star formation appear to have fairly smooth profiles
(e.g., Padoan et al. 1999, Figure 4 for L1448). Broad faint wings are
common (Falgarone \& Phillips 1990).  The ratios of line-wing
intensities for different isotopes of the same molecule typically
vary more than the line-core ratios across the face of a
cloud.  However, the ratio of intensities for transitions from
different levels in the same isotope is approximately constant for
both the cores and the wings (Falgarone et al. 1998; Ingalls et
al. 2000; Falgarone, Pety \& Phillips 2001).

Models have difficulty reproducing all these features. If the
turbulence correlation length is small compared with the photon
mean free path (microturbulence), then the profiles appear
flat-topped or self-absorbed because of non-LTE effects (e.g.,
Liszt et al. 1974; Piehler \& Kegel 1995 and references therein).
If the correlation length is large (macroturbulence), then the
profiles can be Gaussian, but they are also jagged if the number
of correlation lengths is small. Synthetic velocity fields with
steep power spectra give non-Gaussian shapes (Dubinski, Narayan \&
Phillips 1995).

Falgarone et al. (1994) analyzed profiles from a decaying $512^3$
hydrodynamic simulation of transonic turbulence and found line
skewness and wings in good agreement with the Ursa Majoris cloud.
Padoan et al. (1998, 1999) got realistic line profiles from Mach
$5-8$ 3D MHD simulations having super-Alfv\'enic motions with no
gravity, stellar radiation, or outflows. Both groups presented
simulated profiles that were too jagged when the Mach numbers were
high. For example, L1448 in Padoan et al. (1999; Figures 3 and 4) has
smoother $^{13}$CO profiles than the simulations even though this
is a region with star-formation. Large-scale forcing in these
simulations also favors jagged profiles by producing a small
number of strong shocks.

Ossenkopf (2002) found jagged structure in CO line
profiles modelled with $128^3-256^3$ hydrodynamic and MHD
turbulence simulations. He suggested that subgrid velocity
structure is needed to smooth them, but noted that the subgrid
dispersion has to be nearly as large as the total dispersion. The
sonic Mach numbers were very large in these simulations (10--15),
and the forcing was again applied on the largest scales. Ossenkopf
noted that the jaggedness of the profiles could be reduced if the
forcing was applied at smaller scales (producing more shock
compressions along each line of sight), but found that these
models did not match the observed delta-variance scaling.
Ossenkopf also found that subthermal excitation gave line profiles
broad wings without requiring intermittency (Falgarone \& Phillips
1990) or vorticity (Ballesteros-Paredes, Hartmann \& V\'azquez-Semandini 1999), although the
observed line wings seem thermally excited (Falgarone, Pety \&
Phillips 2001).

The importance of unresolved structure in line profiles is
unknown. Falgarone et al. (1998) suggested that profile smoothness
in several local clouds implies emission cells smaller than
$10^{-3}$ pc, and that velocity gradients as large as 16 km
s$^{-1}$ pc$^{-1}$ appear in channel maps. Such gradients were
also inferred by Miesch et al. (1999) based on the large
Taylor-scale Reynolds number for interstellar clouds (this number
measures the ratio of the rms size of the velocity gradients to
the viscous scale, $L_K$, see Section \ref{sect:be}). Tauber et
al. (1991) suggested that CO profiles in parts of Orion were so
smooth that the emission in each beam had to originate in an
extremely large number, $10^6$, of very small clumps, AU-size, if
each clump has a thermal linewidth.  They required $10^4$ clumps
if the internal dispersions are larger, $\sim1$ km s$^{-1}$.
Fragments of $\sim10^{-2}$ pc size were inferred directly from CCS
observations of ragged line profiles in Taurus (Langer et al.
1995).

Recently, Pety \& Falgarone (2003) found small ($<0.02$ pc) regions
with very large velocity gradients in centroid difference maps of
molecular cloud cores. These gradient structures were not
obviously correlated with column density or density, in which case
they would not be shocks.  They could be shear flows, as in the
dissipative regions of subsonic turbulence. Highly supersonic
simulations have apparently not produced such sheared regions yet.
Perhaps supersonic turbulence has this shear in the form of tiny
oblique shocks that simulations cannot yet reproduce with their
high numerical viscosity at the resolution limit.  Alternatively,
ISM turbulence could be mostly decaying, in which case it could be
dominated by low Mach number shocks (Smith, Mac Low \& Heitsch 2000; Smith, Mac Low \& Zuev 2000).

\section{POWER SOURCES FOR INTERSTELLAR TURBULENCE}
\label{sect:heating}

The physical processes by which kinetic energy gets converted into
turbulence are not well understood for the ISM. The main sources
for large-scale motions are: ({\it a}) stars, whose energy input
is in the form of protostellar winds, expanding {\it H II}
regions, O star and Wolf-Rayet winds, supernovae, and combinations
of these producing superbubbles;  ({\it b})  galactic rotation in
the shocks of spiral arms or bars, in the Balbus-Hawley (1991)
instability, and in the gravitational scattering of cloud
complexes at different epicyclic phases; ({\it c})  gaseous
self-gravity through swing-amplified instabilities and cloud
collapse; ({\it d}) Kelvin-Helmholtz and other fluid
instabilities, and ({\it e}) galactic gravity during disk-halo
circulation, the Parker instability, and galaxy interactions.

Sources for the small-scale turbulence observed by radio
scintillation ({\it Interstellar Turbulence II}) include sonic
reflections of shock waves hitting clouds (Ikeuchi \& Spitzer
1984, Ferriere et al. 1988), cosmic ray streaming and other
instabilities (Wentzel 1969b, Hall 1980), field star motions
(Deiss, Just \& Kegel 1990) and winds, and energy cascades from
larger scales (Lazarian, Vishniac \& Cho 2004).  We concentrate on
the large-scale sources here.

Van Buren (1985) estimated that winds from massive main-sequence
stars and Wolf-Rayet stars contribute comparable amounts,
$1\times10^{-25}$ erg cm$^{-3}$ s$^{-1}$, supernovae release about
twice this, and winds from low-mass stars and planetary nebulae
are negligible. Van Buren did not estimate the rate at which this
energy goes into turbulence, which requires multiplication by an
efficiency factor of $\sim0.01-0.1,$ depending on the source. Mac
Low \& Klessen (2004) found that main-sequence winds are
negligible except for the highest-mass stars, in which case
supernovae dominate all the stellar sources, giving
$3\times10^{-26}$ erg cm$^{-3}$ s$^{-1}$ for the energy input,
after multiplying by an efficiency factor of 0.1. Mac Low \&
Klessen (2004) also derived an average injection rate from
protostellar winds equal to $2\times10^{-28}$ erg cm$^{-3}$
s$^{-1}$ including an efficiency factor of $\sim0.05$. {\it H II}
regions are much less important as a general source of motions
because most of the stellar Lyman continuum energy goes into
ionization and heat (Mac Low \& Klessen 2004). Kritsuk \& Norman
(2002a) suggested that moderate turbulence can be maintained by
variations in the background nonionizing UV radiation (Parravano
et al. 2003).

These estimates agree well with the more detailed ``grand source
function'' estimated by Norman \& Ferrara (1996), who also
considered the spatial range for each source. They recognized that
most Type II SNe contribute to cluster winds and superbubbles,
which dominate the energy input on scales of $100-500$ pc (Oey \&
Clarke 1997). Superbubbles are also the most frequent pressure
disturbance for any random disk position (Kornreich \& Scalo
2000).

Power rates for turbulence inside molecular clouds may exceed
these global averages. For example, Stone, Ostriker \& Gammie
(1998) suggested that the turbulent heating rate inside a giant
molecular cloud (GMC) is $\sim1-6\times10^{-27}n_H\Delta v^3/R$
erg cm$^{-3}$ s$^{-1}$ for velocity dispersion $\Delta v$ in km
s$^{-1}$ and size $R$ in pc. For typical $n_H\sim10^2-10^3$
cm$^{-3}$, $\Delta v\sim2$ and $R\sim10$, this exceeds the global
average for the ISM by a factor of $\sim10$, even before internal
star formation begins (see also Basu \& Murali 2001). This
suggests that power density is not independent of scale as it is
in a simple Kolmogorov cascade. An alternative view was expressed
by Falgarone, Hily-Blant \& Levrier (2003) who suggested that the
power density is about the same for the cool and warm phases,
GMCs, and dense cores.  In either case, self-gravity contributes
to the power density locally, and even without self-gravity,
dissipation is intermittent and often concentrated in small
regions.

Galactic rotation has a virtually unlimited supply of energy if it
can be tapped for turbulence (Fleck 1981).  Several mechanisms
have been proposed.  Magneto-rotational instabilities (Sellwood \&
Balbus 1999, Kim et al. 2003) pump energy into gas motion at a
rate comparable to the magnetic energy density times the angular
rotation rate. This was evaluated by Mac Low \& Klessen (2004) to
be $3\times10^{-29}$ erg cm$^{-3}$ s$^{-1}$ for $B=3\mu$G. This is
smaller than the estimated stellar input rate by a factor of
$\sim1000$, but it might be important in the galactic outer
regions where stars form slowly (Sellwood \& Balbus 1999) and in
low-surface brightness galaxies.  Piontek \& Ostriker (2004)
considered how reduced dissipation can enhance the power input to
turbulence from magnetorotational instabilities.

Rotational energy also goes into the gas in spiral shocks where
the fast-moving interspiral medium hits the slower moving dense
gas in a density wave arm (Roberts 1969). Additional input comes
from the gravitational potential energy of the arm as the gas
accelerates toward it. Some of this energy input will be stored in
magnetic compressional energy, some will be converted into
gravitational potential energy above the midplane as the gas
deflects upward (Martos \& Cox 1998), and some will be lost to
heat.  The fraction that goes into turbulence is not known, but the
total power available is
$0.5\rho_{ism}v_{sdw}^3/(2H)\sim5\times10^{-27}$  erg cm$^{-3}$
s$^{-1}$ for interspiral density $\rho_{ism}\sim0.1$ m$_{\rm H}$
cm$^{-3}$, shock speed $v_{sdw}\sim30$ km s$^{-1}$, and half disk
thickness $H=100$ pc. Zhang et al. (2001) suggest that a spiral wave has
driven turbulence in the Carina molecular clouds because the
linewidth-size relation is not correlated with distance from the
obvious sources of stellar energy input.

Fukunaga \& Tosa (1989) proposed that rotational energy goes to
clouds that gravitationally scatter off each other during random
phases in their epicycles. Gammie et al. (1991) estimated that the
cloud velocity dispersion can reach the observed value of $\sim5$
km s$^{-1}$ in this way.  Vollmer \& Beckert (2002) considered the
same mechanism with shorter cloud lifetimes and produced a steady
state model of disk accretion. A second paper (Vollmer \& Beckert
2003) included supernovae.

The gravitational binding energy in a galaxy disk heats the
stellar population during swing-amplified shear instabilities that
make flocculent spiral arms (e.g., Fuchs \& von Linden 1998). It
can also heat the gas (Thomasson, Donner \& Elmegreen 1991; Bertin
\& Lodato 2001; Gammie 2001) and feed turbulence (Huber \&
Pfenniger 2001; Wada, Meurer \& Norman 2002). Continued collapse
of the gas may feed more turbulence on smaller scales (Semelin et
al. 1999, Chavanis 2002, Huber \& Pfenniger 2002). A gravitational
source of turbulence is consistent with the observed power spectra
of flocculent spiral arms (Elmegreen et al. 2003). The energy
input rate for the first e-folding time of the instability is
approximately the ISM energy density, $1.5\rho \Delta v^2$, times
the growth rate $2\pi G\rho H/\Delta v$ for velocity dispersion
$\Delta v$. This is $\sim10^{-27}$ erg cm$^{-3}$ s$^{-1}$ in the
Solar neighborhood---less than supernovae by an order of
magnitude. However, continued energy input during cloud collapse
would increase the power available for turbulence in proportion to
$\rho^{4/3}$. The efficiency for the conversion of gravitational
binding energy into turbulence is unknown, but because
gravitational forces act on all of the matter and, unlike stellar
explosions, do not require a hot phase of evolution during which
energy can radiate, the efficiency might be high.

Conventional fluid instabilities provide other sources of
turbulence on the scales over which they act. For example, a cloud
hit by a shock front will shed its outer layers and become
turbulent downstream (Xu \& Stone 1995), and the interior of the
cloud can be energized as well (Miesch \& Zweibel 1994, Kornreich
\& Scalo 2000). Cold decelerating shells have a kinematic
instability (Vishniac 1994) that can generate turbulence inside
the swept-up gas (Blondin \& Marks 1996, Walder \& Folini 1998).
Bending mode and other instabilities in cloud collisions generate
a complex filamentary structure (Klein \& Woods 1998). It is also
possible that the kinetic energy of a shock can be directly
converted into turbulent energy behind the shock (Rotman 1991;
Andreopoulos, Agui \& Briassulis 2000). Kritsuk \& Norman
(2002a,b) discuss how thermal instabilities can drive turbulence,
in which case the underlying power source is stellar radiation
rather than kinetic energy. There are many individual sources for
turbulence, but the energy usually comes from one of the main
categories of sources listed above.

Sources of interstellar turbulence span such a wide range of
scales that it is often difficult to identify any particular
source for a given cloud or region. Little is known about the
behavior of turbulence that is driven like this. The direction and
degree of energy transfer and the morphology of the resulting flow
could be greatly affected by the type and scale of energy input
(see Biferale et al. 2004).  However, it appears that for average
disk conditions the power input is dominated by cluster winds or
superbubbles with an injection scale of $\sim50-500$ pc.

\section{THEORY OF INTERSTELLAR TURBULENCE}

\subsection{What is Turbulence and Why Is It So Complicated?}
\label{sect:complicated}

Turbulence is nonlinear fluid motion resulting in the excitation
of an extreme range of correlated spatial and temporal scales.
There is no clear scale separation for perturbation
approximations, and the number of degrees of freedom is too large
to treat as chaotic and too small to treat in a statistical
mechanical sense. Turbulence is deterministic and unpredictable,
but it is not reducible to a low-dimensional system and so does
not exhibit the properties of classical chaotic dynamical systems.
The strong correlations and lack of scale separation preclude the
truncation of statistical equations at any order. This means that
the moments of the fluctuating fields evaluated at high order
cannot be interpreted as analogous to moments of the microscopic
particle distribution, i.e., the rms velocity cannot be used as a
pressure.

Hydrodynamic turbulence arises because the nonlinear advection
operator, $({\bf u}\cdot\nabla){\bf u}$, generates severe
distortions of the velocity field by stretching, folding, and
dilating fluid elements.  The effect can be viewed as a continuous
set of topological deformations of the velocity field (Ottino
1989), but in a much higher dimensional space than chaotic systems
so that the velocity field is, in effect, a stochastic field of
nonlinear straining. These distortions self-interact to generate
large amplitude structure covering the available range of scales.
For incompressible turbulence driven at large scales, this range
is called the inertial range because the advection term
corresponds to inertia in the equation of motion. For a purely
hydrodynamic incompressible system, this range is measured by the
ratio of the advection term to the viscous term, which is the
Reynolds number $Re = UL/\nu\sim3\times10^3M_aL_{pc}n$, where $U$
and $L$ are the characteristic large-scale velocity and length,
$L_{pc}$ is the length in parsecs, $M_a$ is the Mach number, $n$
is the density, and $\nu$ is the kinematic viscosity. In the cool
ISM, $Re\sim10^5$ to $10^7$ if viscosity is the damping mechanism
(less if ambipolar diffusion dominates; Section \ref{sect:sim}).
Another physically important range is the Taylor scale Reynolds
number, which is $Re_\lambda=U_{rms}L_T/\nu$ for $L_T=$ the ratio
of the rms velocity to the rms velocity gradient (see Miesch et
al. 1999).

With compressibility, magnetic fields, or self-gravity, all the
associated fields are distorted by the velocity field and exert
feedback on it. Hence, one can have MHD turbulence, gravitational
turbulence, or thermally driven turbulence, but they are all
fundamentally tied to the advection operator. These additional
effects introduce new globally conserved quadratic quantities and
eliminate others (e.g., kinetic energy is not an inviscid conserved
quantity in compressible turbulence), leading to fundamental
changes in the behavior. This may affect the way energy is
distributed among scales, which is often referred to as the
cascade.

Wave turbulence occurs in systems dominated by nonlinear wave
interactions, including plasma waves (Tsytovich 1972), inertial
waves in rotating fluids (Galtier 2003), acoustic turbulence
(L'vov, L'vov \& Pomyalov 2000), and internal gravity waves
(Lelong \& Riley 1991). A standard procedure for treating these
weakly nonlinear systems is with a kinetic equation that describes
the energy transfer attributable to interactions of three (in some
cases four) waves with conservation of energy and momentum (see
Zakharov, L'vov \& Falkovich 1992). Wave turbulence is usually
propagating, long-lived, coherent, weakly nonlinear, and weakly
dissipative, whereas fully developed fluid turbulence is
diffusive, short-lived, incoherent, strongly nonlinear, and
strongly dissipative (Dewan 1985).  In both wave and fluid
turbulence, energy is transferred among scales, and when it is fed
at the largest scales with dissipation at the smallest scales, a
Kolmogorov or other power-law power spectrum often results. Court
(1965) suggested that wave turbulence be called undulence to
distinguish it from the different physical processes involved in
fully developed fluid turbulence.

\subsection{Basic Equations}
\label{sect:be}

The equations of mass and momentum conservation are
\begin{equation}
\partial \rho/\partial t + \nabla\cdot\left(\rho{\bf u}\right)=
0,\end{equation}
\begin{equation}
\partial {\bf u}/\partial t + \left({\bf u}\cdot\nabla\right){\bf u}
= -{1\over{\rho}}\nabla P+{\bf F}+{1\over{\rho}}\nabla\cdot{\bf
\sigma},\label{eq:ns}\end{equation} where $\rho$, ${\bf u}$, $P$
are the mass density, velocity, and pressure, and ${\bf \sigma}$
is the shear stress tensor.  The last term is often written as
$\nu\left(\nabla^2{\bf u}+\nabla\left[\nabla\cdot{\bf
u}\right]/3\right)$, where $\nu\sim10^{20}c_5/n$ is the kinematic
viscosity (in cm$^2$ s$^{-1}$) for thermal speed $c_5$ in units of
$10^5$ cm s$^{-1}$ and density $n$ in cm$^{-3}$.  This form of
${\bf \sigma}$ is only valid in an incompressible fluid since the
viscosity depends on density (and temperature). The scale $L_K$ at
which the dissipation rate equals the advection rate is called the
Kolmogorov microscale and is approximately
$L_K=10^{15}/\left(nM_a\right)$ cm.

In the compressible case, the scale at which dissipation dominates
advection will vary with position because of the large density
variations. This could spread out the region in wave number space
at which any power-law cascade steepens into the dissipation
range. Usually the viscosity is assumed to be constant for ISM
turbulence. The force per unit mass ${\bf F}$ may include
self-gravity and magnetism, which introduce other equations to be
solved, such as the Poisson and induction equations. The pressure
$P$ is related to the other variables through the internal energy
equation. For this discussion, we assume an isentropic (or
barotropic) equation of state, $P\sim\rho^\gamma$, with $\gamma$ a
parameter ($ = 1$ for an isothermal gas). The inclusion of an
energy equation is crucial for ISM turbulence; otherwise the
energy transfer between kinetic and thermal modes may be
incorrect. An important dimensionless number is the sonic Mach
number $M_a=u/a$, where $a$ is the sound speed. For most ISM
turbulence, $M_a\sim0.1$ - 10, so it is rarely incompressible and
often supersonic, producing shocks.

\subsection{Statistical Closure Theories}

In turbulent flows all of the variables in the hydrodynamic
equations are strongly fluctuating and can be described only
statistically.  The traditional practice in incompressible
turbulence is to derive equations for the two- and three-point
correlations as well as various other ensemble averages. An
equation for the three-point correlations generates terms
involving four-point correlations, and so on. The existence of
unknown high-order correlations is a classical closure problem,
and there are a large number of attempts to close the equations,
i.e., to express the high-order correlations in terms of the
low-order correlations.  Examples range from simple gradient
closures for mean flow quantities to mathematically complex
approaches such as the Lagrangian history Direct Interaction
Approximation (DIA, see Leslie 1973), the Eddy-Damped Quasi-Normal
Markovian (EDQNM) closure (see Lesieur 1990), diagrammatic
perturbation closures, renormalization group closures, and others
(see general references given in Section \ref{sect:intro}).

The huge number of additional unclosed terms that are generated by
compressible modes and thermal modes using an energy equation (see
Lele 1994) render conventional closure techniques ineffective for
much of the ISM, except perhaps for extremely small Mach numbers
(e.g., Bertoglio, Bataille \& Marion 2001). Besides their
intractability, closure techniques only give information about the
correlation function or power spectrum, which yields an incomplete
description because all phase information is lost and higher-order
moments are not treated.  An exact equation for the infinite-point
correlation functional can be derived (Beran 1968) but not solved.
For these reasons we do not discuss closure models here.  Other
theories involving scaling arguments (e.g., She \& Leveque 1994),
statistical mechanical formulations (e.g., Shivamoggi 1997), shell
models (see Biferale 2003 for a review), and dynamical
phenomenology (e.g., Leorat et al. 1990) may be more useful, along
with closure techniques for the one-point pdfs, such as the
mapping closure (Chen, Chen \& Kraichnan 1989).

\subsection{Solenoidal and Compressible Modes}
\label{sect:sol}

For compressible flows relevant to the ISM, the Helmholtz
decomposition theorem splits the velocity field into compressible
(dilatational, longitudinal) and solenoidal (rotational, vortical)
modes $u_c$ and $u_s$, defined by $\nabla\times{\bf u}_c = 0$ and
$\nabla\cdot {\bf u}_s = 0$. In strong 3D turbulence these
components have different effects, leading to shocks and
rarefactions for ${\bf u}_c$ and to vortex structures for ${\bf
u}_s$. Only the compressible mode is directly coupled to the
gravitational field. The two modes are themselves coupled and
exchange energy. The coupled evolution equations for the vorticity
$\omega=\nabla\times{\bf u}$ and dilatation $\nabla\cdot{\bf u}$
contain one asymmetry that favors transfer from solenoidal to
compressible components in the absence of viscous and pressure
terms (V\'azquez-Semadeni, Passot \& Pouquet 1996; Kornreich \&
Scalo 2000) and another asymmetry that transfers from compressible
to solenoidal when pressure and density gradients are not aligned,
as in an oblique shock. This term in the vorticity equation,
proportional to $\nabla P \times \nabla(1/\rho)$, causes
baroclinic vorticity generation. If turbulence is modeled as
barotropic or isothermal, vorticity generation is suppressed. One
or the other of these asymmetries can dominate in different parts
of the ISM (Section \ref{sect:sim}).

Only the solenoidal mode exists in incompressible turbulence, so
vortex models can capture much of the dynamics (Pullin \& Saffman
1998). Compressible supersonic turbulence has no such conceptual
simplification, because even at moderate Mach numbers there will
be strong interactions with the solenoidal modes, and with the
thermal modes if isothermality is not assumed.

\subsection{Global Inviscid Quadratic Invariants are Fundamental
Constraints on the Nature of Turbulent Flows} \label{sect:global}

Quadratic-conserved quantities constrain the evolution of
classical systems.  A review of continuum systems with dual
quadratic invariants is given by Hasegawa (1985). For turbulence,
the important quadratic invariants are those conserved by the
inviscid momentum equation. The conservation properties of
turbulence differ for incompressible versus compressible, 2D
versus 3D, and hydrodynamic versus MHD (see also Biskamp 2003).
For incompressible flows, the momentum equation neglecting
viscosity and external forces is, from Equation \ref{eq:ns} above,
\begin{equation}
\partial{\bf u}/\partial t + \left({\bf u}\cdot\nabla\right){\bf u}=-\nabla P/\rho.
\end{equation}
Taking the scalar product of this equation with {\bf u},
integrating over all space, using Gauss' theorem, and assuming
that the velocity and pressure terms vanish at infinity gives
\begin{equation}
\partial/\partial t\int\left(\onehalf u^2\right){\bf
dr}=0\end{equation} demonstrating that kinetic energy per unit
mass is globally conserved by the advection operator.  In Fourier
space this equation leads to a ``detailed'' conservation condition
in which only triads of wavevectors participate in energy
transfer. It is this property that makes closure descriptions in
Fourier space so valuable. Virtually all the phenomenology
associated with incompressible turbulence traces back to the
inviscid global conservation of kinetic energy; unfortunately,
compressible turbulence does not share this property. Another
quadratic conserved quantity for incompressible turbulence is
kinetic helicity $<{\bf u}\cdot\omega>$, which measures the
asymmetry between vorticity and velocity fields. This is not
positive definite so its role in constraining turbulent flows is
uncertain (see Chen, Chen \& Eyink 2003). Kurien, Taylor \&
Matsumoto (2003) find that helicity conservation controls the
inertial range cascade at large wave numbers for incompressible
turbulence.

For 2D turbulence there is an additional positive-definite
quadratic invariant, the enstrophy or mean square vorticity
$<\left(\nabla \times{\bf u}\right)^2>$.  The derivative in the
enstrophy implies that it selectively decays to smaller scales
faster than the energy, with the result that the kinetic energy
undergoes an inverse cascade to smaller wave numbers.  This result
was first pointed out by Kraichnan (1967) and has been verified in
numerous experiments and simulations (see Tabeling 2002 for a
comprehensive review of 2D turbulence). Because of the absence of
the vorticity stretching term $\omega\cdot\nabla u$ in the 2D
vorticity equation, which produces the complex system of vortex
worms seen in 3D turbulence, 2D turbulence evolves into a system
of vortices that grow with time through their merging.

In 2D-compressible turbulence the square of the potential
vorticity ($\omega/\rho$) is also conserved (Passot, Pouquet \&
Woodward 1988), unlike the case in 3D-compressible turbulence.
This means that results obtained from 2D simulations may not apply
to 3D. However, if energy is fed into galactic turbulence at very
large scales, e.g., by galactic rotation, then the ISM may
possesses quasi-2D structure on these scales.

Compressibility alters the conservation properties of the flow.
Momentum density $\rho{\bf u}$ is the only quadratic-conserved
quantity but it is not positive definite. Momentum conservation
controls the properties of shocks in isothermal supersonic
nonmagnetic turbulence.  Interacting oblique shocks generate flows
that create shocks on smaller scales (Mac Low \& Norman 1993),
possibly leading to a shock cascade controlled by momentum
conservation. Kinetic energy conservation is lost because it can
be exchanged with the thermal energy mode: compression and shocks
heat the gas. The absence of kinetic energy conservation means
that in Fourier space the property of detailed conservation by
triads is lost, and so is the basis for many varieties of
statistical closures.   Energy can also be transferred between
solenoidal and compressional modes, and even when expanded to only
second order in fluctuation amplitude, combinations of any two of
the vorticity, acoustic, and entropy modes, or their
self-interactions, generate other modes.

The assumption of isothermal or isentropic flow forces a helicity
conservation that is unphysical in general compressible flows; it
suppresses baroclinic vorticity creation, and affects the exchange
between compressible kinetic and thermal modes. Isothermality is
likely to be valid in the ISM only over a narrow range of
densities, from $10^3$ to $10^4$ cm$^{-3}$ in molecular clouds for
example (Scalo et al. 1998), and this is much smaller than the
density variation found in simulations. Isothermality also
suppresses the ability of sound waves to steepen into shocks.

Magnetic fields further complicate the situation. Energy can be
transferred between kinetic and magnetic energy so only the total,
$\onehalf\rho u^2 + B^2/8\pi$, is conserved. Kolmogorov-type
scaling may still result in the cross-field direction (Goldreich
\& Sridhar 1995). The magnetic helicity ${\bf A}\cdot{\bf B}$
(${\bf A}=$ magnetic vector potential) and cross helicity ${\bf
u}\cdot{\bf B}$ are conserved but their role in the dynamics is
uncertain. Pouquet et al. (1976) addressed the MHD energy cascade
using EDQNM closure and predicted an inverse cascade for large
magnetic helicities. Ayyer \& Verma (2003) assumed a Kolmogorov
power spectrum and found that the cascade direction depends on
whether kinetic and magnetic helicities are nonzero. Without
helicity, all the self- and cross-energy transfer between kinetic
and magnetic energies are direct, meaning to larger wave numbers,
whereas the magnetic and kinetic helical contributions give an
inverse cascade for the four sets of possible magnetic and kinetic
energy exchanges, confirming the result of Pouquet et al. (1976).
With or without helicity, the net energy transfer within a given
wave number band is from magnetic to kinetic on average.  It is
seen that while kinetic helicity cannot reverse the direct cascade
in hydrodynamic turbulence, magnetic helicity in MHD turbulence
can.  The effects of kinetic and magnetic helicity on supersonic
hydrodynamic or MHD turbulence are unexplored.

These considerations suggest that the type of cascade expected for
highly compressible MHD turbulence in the ISM cannot be predicted
yet. Numerical simulations may result in an incorrect picture of
the cascade if restricted assumptions like isothermality and
constant kinematic viscosity are imposed.

\subsection{Scale-invariant Kinetic Energy Flux and Scaling
Arguments for the Power Spectrum} \label{sect:scale}

In the Kolmogorov (1941) model of turbulence energy injected at
large scales ``cascades'' to smaller scales by
kinetic-energy--conserving interactions that are local in Fourier
space at a rate independent of scale, and then dissipates by
viscosity at a much smaller scale.  At a given scale $\ell$ having
velocity $u_\ell$, the rate of change of kinetic energy per unit
mass from exchanges with other scales is $u_\ell^2$ divided by the
characteristic timescale, taken to be $\ell/u_\ell$. The constant
energy rate with scale gives $u_\ell\sim\ell^{1/3}$. If one
considers $u_\ell$ to represent the ensemble average velocity
difference over separation $\ell$, this result gives the
second-order structure function $S_2(\ell)\sim \ell^{2/3}$. In
terms of wave number, $u_k^2\sim k^{-2/3}$, so the kinetic energy
per unit mass per unit wave number, which is the energy spectrum
$E(k)$ (equal to $4\pi k^2 P(k)$ where $P(k)$ is the 3D power
spectrum), is given by
\begin{equation}E(k)\sim u_k^2/k \propto k^{-5/3}.\end{equation}
This 5/3 law was first confirmed experimentally 20 years after
Kolmogorov's proposal using tidal channel data to obtain a large
inertial range (Grant, Stewart \& Moillet 1962).  More recent
confirmations of this relation and the associated second-order
structure function have used wind tunnels, jets, and
counter-rotating cylinders (see Frisch 1995). The largest 3D
incompressible simulations (Kaneda et al. 2003) indicate that the
power-law slope of the energy spectrum might be steeper by about
0.1. Similar considerations applied to the helicity cascade
(Kurien et al. 2003) predict a transition to $E(k)\sim k^{-4/3}$
at large $k$, perhaps explaining the flattening of the spectrum,
or ``bottleneck effect'' seen in many simulations. Leorat, Passot
\& Pouquet (1990) generalized the energy cascade phenomenology to
the compressible case assuming locality of energy transfer (no
shocks), but allowing for the fact that the kinetic energy
transfer rate will not be constant. It is commonly supposed that
highly supersonic turbulence should have the spectrum of a field
of uncorrelated shocks, $k^{-2}$ (Saffman 1971), although this
result has never been derived in more than one-dimension, and
shocks in most turbulence should be correlated.

These scaling arguments make only a weak connection with the
hydrodynamical equations, either through a conservation property
or an assumed geometry (see also Section \ref{sect:complicated}).
The only derivation of the $-5/3$ spectrum for incompressible
turbulence that is based on an approximate solution of the
Navier-Stokes equation is Lundgren's (1982) analysis of an
unsteady stretched-spiral vortex model for the fine-scale
structure (see Pullin \& Saffman 1998). No such derivation exists
for compressible turbulence.

Kolmogorov's (1941) most general result is the ``four-fifths
law.'' For statistical homogeneity, isotropy, and a stationary
state driven at large scales, the energy flux through a given
wave number band $k$ should be independent of $k$ for
incompressible turbulence (Frisch 1995, Section 6.2.4). Combined with
a rigorous relation between energy flux and the third-order
structure function, this gives (Frisch 1995) an exact result in
the limit of small $\delta r$:
\begin{equation} S_3(\delta r) = < \left(\delta v(r,\delta
r\right)^3> = -(4/5) \epsilon \delta r;\end{equation}$\epsilon$ is
the rate of dissipation due to viscosity, $\delta v\left(r,\delta
r\right)$ is the velocity difference over spatial lag $\delta r$
at position $r$, and the ensemble average is over all positions.
Kolmogorov (1941) obtained this result for decaying turbulence
using simpler arguments. Notice that self-similarity is not
assumed.  A more general version of this relation has been
experimentally tested by Moisy, Tabeling, \& Willaime (1999) and
found to agree to within a few percent over a range of scales up
to three decades.

For compressible ISM turbulence, kinetic energy is not conserved
between scales; for this reason, the term ``inertial range'' is
meaningless.  As a result, there is no guarantee of scale-free or
self-similar power-law behavior. The driving agents span a wide
range of scales (Section \ref{sect:heating}) and the ISM is larger
than all of them. Excitations may spread to both large and small
scales, independent of the direction of net energy transfer.  In
the main disks of galaxies where the Toomre parameter $Q$ is less
than $\sim2$, vortices larger than the disk thickness may result
from quasi-2D turbulence; recall that $Q$ equals the ratio of the
scale height to the epicyclic length. Large-scale vortical motions
in a compressible, self-gravitating ISM may amplify to look like
flocculent spirals.

\subsection{Intermittency and Structure Function Scaling}
\label{sect:interm}

Kolmogorov's (1941) theory did not recognize that dissipation in
turbulence is ``intermittent'' at small scales, with intense
regions of small filling factor, giving fat, nearly exponential
tails in the velocity difference or other probability distribution
functions (pdfs). Intermittency can refer to either the time,
space, or probability structures that arise. The best-studied
manifestation is ``anomalous scaling'' of the high-order velocity
structure functions (Equation \ref{eq:sf}), $S(\delta
r)=\left(v[r]-v[r+\delta r]\right)^p\sim \delta r^{\zeta_p}$. For
Kolmogorov turbulence with the four-fifths law and an additional
assumption of self-similarity ($\delta v(\lambda \delta
r)=\lambda^h\delta v(\delta r)$ for small $\delta r$ with $h=1/3$)
$\zeta_p=p/3$.  In real incompressible turbulence, $\zeta_p$ rises
more slowly with $p$ (compare Section \ref{sect:goldreich}; e.g.,
Anselmet et al. 1984).

Interstellar turbulence is probably intermittent, as indicated by
the small filling fraction of clouds and their relatively high
energy dissipation rates (Section \ref{sect:heating}).  Velocity
pdfs and velocity difference pdfs also have fat tails (Section
\ref{sect:diag}), as may elemental abundance distributions ({\it
Interstellar Turbulence II}). Other evidence for intermittency in
the dissipation field is the $10^3$ K collisionally excited gas
required to explain CH+, HCO+, OH, and excited H$_2$ rotational
lines (Falgarone et al. 2004; {\it Interstellar Turbulence II}).
Although it is difficult to distinguish between the possible
dissipation mechanisms (Pety \& Falgarone 2000), the dissipation
regions occupy only $\sim1$\% of the line of sight and they seem
to be ubiquitous (Falgarone et al. 2004). If they are viscous
shear layers, then their sizes ($10^{15}$ cm) cannot be resolved
with present-day simulations.

\begin{figure}
\plotone{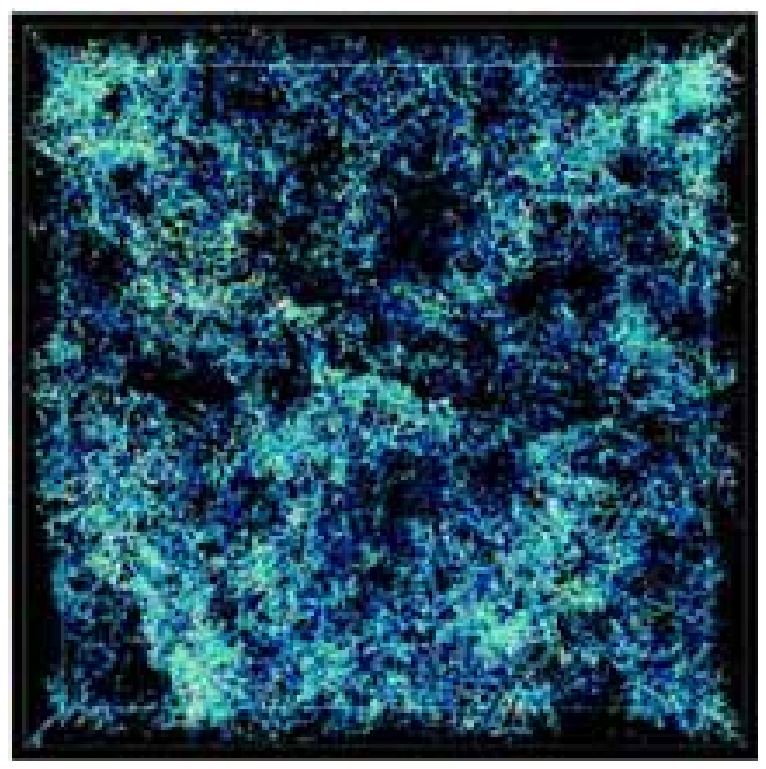} 
\caption{Vorticity structures in
a $1024\times1024\times128$ section of a $1024^3$ simulation of
compressible decaying nonmagnetic turbulence with initial rms Mach
number of unity. From Porter, Woodward \& Pouquet
(1998).{\it (for slightly higher resolution, see 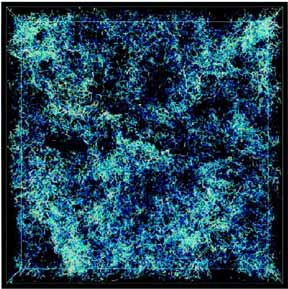)}}\label{fig:porter}\end{figure}

A geometrical model for $\zeta_p$ in incompressible nonmagnetic
turbulence was proposed by She \& Leveque (1994, see also Liu \&
She 2003). The model considers a box of turbulent fluid
hierarchically divided into sub-boxes in which the energy
dissipation is either large or small.  The mean energy flux is
conserved at all levels with Kolmogorov scaling, and the
dissipation regions are assumed to be one-dimensional vortex tubes
or worms, known from earlier experiments and simulations. Then
$\zeta_p$ was derived to be $p/9 + 2(1-[2/3]^{p/3})$ and shown to
match the experiments up to at least $p=10$ (see Dubrulle 1994 and
Boldyrev 2002 for derivations). Porter, Pouquet \& Woodward (2002)
found a flow dominated by vortex tubes in simulations of decaying
transonic turbulence at fairly small Mach numbers (Figure
\ref{fig:porter}) and got good agreement with the She-Leveque
formula. A possible problem with the She-Leveque approach is that
the dimension of the most intense vorticity structures does not
have a single value, and the average value is larger than unity
for incompressible turbulence (Vainshtein 2003). A derivation of
$\zeta_p$ based on a dynamical vortex model has been given by
Hatakeyama \& Kambe (1997).

Politano \& Pouquet (1995) proposed a generalization for the MHD
case (see also \ref{sect:goldreich}). It depends on the scaling
relations for the velocity and cascade rate, and on the
dimensionality of the dissipative structures. They noted solar
wind observations that suggested the most dissipative structures
are two dimensional current sheets. M\"uller \& Biskamp (2000)
confirmed that dissipation in incompressible MHD turbulence occurs
in 2D structures, in which case $\zeta_p = p/9 + 1 - (1/3)^{p/3}$.
However, Cho, Lazarian \& Vishniac (2002a) found that for
anisotropic incompressible MHD turbulence measured with respect to
the {local} field, She-Leveque scaling with 1D intermittent
structures occurs for the velocity and a slightly different
scaling occurs for the field. Boldyrev (2002) assumed that
turbulence is mostly solenoidal with Kolmogorov scaling, while the
most dissipative structures are shocks, again getting $\zeta_p =
p/9 + 1 - (1/3)^{p/3}$ because of the planar geometry; the
predicted energy spectrum was $E(k)\sim k^{-1-\zeta_2}\sim
k^{-1.74}$. Boldyrev noted that the spectrum will be steeper if
the shocks have a dimension equal to the fractal dimension of the
density.

The uncertainties in measuring ISM velocities preclude a test of
these relations, although they can be compared with numerical
simulations.  Boldyrev, Nordlund \& Padoan (2002b) found good
agreement in 3D super-Alfv\'enic isothermal simulations for both
the structure functions to high order and the power spectrum. The
simulations were forced solenoidally at large scales and have
mostly solenoidal energy, so they satisfy the assumptions in
Boldyrev (2002). Padoan et al. (2003b) showed that the dimension
of the most dissipative structures varies with Mach number from
one for lines at subsonic turbulence to two for sheets at Mach 10.
The low Mach number result is consistent with the nonmagnetic
transonic simulation in Porter et al. (2002; see Figure
\ref{fig:porter}).  Kritsuk \& Norman (2003) showed how
nonisothermality leads to more complex folded structures with
dimensions larger than two.

\subsection{Details of the Energy Cascade: Isotropy and
Independence of Large and Small Scales}

Kolmogorov's model implies an independence between large and small
scales and a resulting isotropy on the smallest scales. Various
types of evidence for and against this prediction were summarized
by Yeung \& Brasseur (1991). One clue comes from the
incompressible Navier-Stokes equation written in terms of the
Fourier-transformed velocity. Global kinetic energy conservation
is then seen to occur only for interactions between triads of
wavevectors (e.g., Section \ref{sect:powerspectra}). The trace of
the equation for the energy spectrum tensor gives the rate of
change of energy per unit wave number in Fourier modes $E(k)$
owing to the exchange of energy with all other modes $T(k)$ and
the loss from viscous dissipation: $\partial E(k)/
\partial t = T(k) - 2\nu k^2 E(k)$.  The details of the
energy transfer can be studied by decomposing the transfer
function T(k) into a sum of contributions $T(k| p,q)$ defined as
the energy transfer to $k$ resulting from interactions between
wave numbers $p$ and $q$.

Domaradzki \& Rogallo (1990) and Yeung \& Brasseur (1991) analyzed
$T(k| p,q)$ from simulations to show that, whereas there is a net
local transfer to higher wave numbers, at smaller and smaller
scales the cascade becomes progressively dominated by nonlocal
triads in which one leg is in the energy-containing (low-$k$)
range.  This means that small scales are not decoupled from large
scales. Waleffe (1992) pointed out that highly nonlocal triad
groups tend to cancel each other in the net energy transfer for
isotropic turbulence; the greatest contribution is from triads
with a scale disparity of about an order of magnitude (Zhou 1993).
Yeung \& Brasseur (1991) and Zhou, Yeung \& Brasseur (1996)
demonstrated that long-range couplings are important in causing
small-scale anisotropy in response to large-scale anisotropic
forcing, and that this effect increases with Reynolds number.

The kinetic energy transfer between scales can be much more
complex in the highly compressible case where the utility of triad
interactions is lost.  Fluctuations with any number of wavevector
combinations can contribute to the energy transfer. Momentum
density is still conserved in triads, but the consequences of this
are unknown. Even in the case of very weakly compressible
turbulence, Bataille \& Zhou (1999) found 17 separate
contributions to the total compressible transfer function $T(k)$.
Nevertheless, the EDQNM closure theory applied to very weakly
compressible turbulence by Bataille \& Zhou (1999) and Bertoglio,
Bataille \& Marion (2001), assuming only triadic interactions,
yields interesting results that may be relevant to warm {\it H I},
the hot ionized ISM, or dense molecular cores where the turbulent
Mach number might be small. Then the dominant compressible energy
transfer is not cascade-type but a cross-transfer involving local
(in spectral space) transfer from solenoidal to compressible
energy.  At Mach numbers approaching unity (for which the theory
is not really valid) the transfer changes to cascade-type, with a
net flux to higher wave numbers.

Pouquet et al. (1976) were among the first to address the locality
and direction of the MHD energy cascade using a closure method.
For more discussions see Ayyer \& Verma (2003), Biferale (2003)
and Biskamp (2003).

\subsection{Velocity Probability Distribution}
\label{sect:velo}

A large number of papers have demonstrated non-Gaussian behavior
in the pdfs of vorticity, energy dissipation, passive scalars, and
fluctuations in the pressure and nonlinear advection, all for
incompressible turbulence. Often the pdfs have excess tails that
tend toward exponentials at small scale (see Chen et al. 1989;
Castaing, Gagne \& Hopfinger 1990 for early references). Velocity
fluctuation {differences} and {derivatives} exhibit tails
of the form $\exp(-v^\alpha)$ with $1<\alpha<2$ and $\alpha\sim1$
at small scale (e.g., Anselmet et al. 1984, She et al. 1993).
Physically this is a manifestation of intermittency, with the most
intense turbulence becoming less space-filling at smaller scales.
The behavior may result from the stretching properties of the
advection operator (see She 1991 for a review).

This same kind of behavior has been observed for the ISM in the
form of excess CO line wings (Falgarone \& Phillips 1990) and in
the pdfs of CO (Padoan et al. 1997, Lis et al. 1998, Miesch et al.
1999) and {\it H I} (Miville-Desch${\hat {\rm e}}$nes, Joncas \&
Falgarone 1999) centroid velocities. The regions include quiescent
and active star formation, diffuse {\it H I}, and self-gravitating
molecular clouds. Lis et al. (1996, 1998) compared the
observations with simulations of mildly supersonic decaying
turbulence and associated the velocity-difference pdf tails with
filamentary structures and regions of large vorticity.

The first comparison of observations with simulations for velocity
difference pdfs was given by Falgarone et al. (1994), who used
optically thin line profile shapes from a simulation of decaying
transonic hydrodynamic turbulence. They found that the excess
wings could be identified in some cases with localized regions of
intense vorticity.  Klessen (2000) compared observations with the
centroid-velocity-difference distributions from isothermal
hydrodynamic simulations and found fair agreement with an approach
to exponential tails on the smallest scales. Smith, Mac Low \&
Zuev (2000) found an exponential distribution of velocity
differences across shocks in simulations of hypersonic decaying
turbulence and were able to derive this result using an extension
of the mapping closure technique (e.g., Gotoh \& Kraichnan 1993).
Ossenkopf \& Mac Low (2002) gave a detailed comparison between
observations of the Polaris flare and MHD simulation velocity and
velocity centroid-difference pdfs, along with other descriptors of
the velocity field. Cosmological-scale galaxy velocity-difference
pdfs (Seto \& Yokoyama 1998) also exhibit exponential forms at
small separations.

The pdf of the velocity itself is not yet understood. The argument
that the velocity pdf should be Gaussian based on the
central-limit theorem applied to independent Fourier coefficients
of an expansion of the velocity field, or to sums of independent
velocity changes (e.g., Tennekes \& Lumley 1972), neglects
correlations and only applies to velocities within a few standard
deviations of the mean. Perhaps this is why the early work found
Gaussian velocity pdfs (e.g., Monin \& Yaglom 1975, Kida \&
Murakami 1989, Figure 6; Jayesh \& Warhaft 1991, Figure 1; Chen et
al. 1993, Figure 3). The velocity pdf must possess nonzero
skewness (unlike a Gaussian) in order to have energy transfer
among scales (e.g., Lesieur 1990).

Recent evidence for non-Gaussian velocity pdfs have been found for
incompressible turbulence from experimental atmospheric data
(Sreenivasan \& Dhruva 1998), turbulent jets (Noullez et al.
1997), boundary layers (Mouri et al. 2003), and quasi-2D
turbulence (Bracco et al. 2000). The sub-Gaussian results (with
pdf flatness factor $F=<u^4>/<u^2>$ less than 3) are probably from
the dominance of a small number of large-scale modes, while the
hyper-Gaussian results ($F>3$) may be the result of correlations
between Fourier modes Mouri et al. 2003), an alignment of vortex
tubes (Takaoka 1995), or intermittency (Schlichting \& Gersten
2000). The results are unexplained quantitatively.

For the ISM, exponential or $1/v$ centroid-velocity distributions
were discovered and rediscovered several times over the past few
decades (see Miesch, Scalo \& Bally 1999). Miesch \& Scalo (1995)
and Miesch et al. (1999) found near-exponential tails in the
$^{13}$CO centroid-velocity pdfs of several molecular regions. The
pdf of the {\it H I}  gas centroid-velocity component
perpendicular to the disk of the LMC is also an exponential (Kim
et al. 1998).

Rigorous derivations of the equation for the velocity pdf of
incompressible hydrodynamic turbulence are presented in Chapter 12
and Appendix H of Pope (2000). Dopazo, Valino \& Fueyo (1997) show
how to derive evolution equations for the moments of the velocity
distribution from the kinetic equation. Closure methods for pdf
equations are discussed in Chen \& Kollman (1994), Dopazo (1994),
Dopazo et al. (1997), and Pope (2000). Results generally predict
Gaussian pdfs, although this depends somewhat on the closure
method and assumptions.

Simulations have not generally examined the velocity pdf in
detail, and when they have, the centroid-velocity pdf or optically
thin line profile is usually given. An important exception is the
3D incompressible simulation of homogeneous shear flows by Pumir
(1996), who found nearly exponential velocity fluctuation pdfs for
velocity components perpendicular to the streamwise component.
However, for conditions more applicable to the compressible ISM,
the results are mixed. Smith, Mac Low \& Zuev (2000) found a
Gaussian distribution of shock speeds in 3D simulations of
hypersonic decaying turbulence, but did not relate this to the pdf
of the total velocity field. Lis et al. (1996) found the pdf of
centroid velocities to be Gaussian or sub-Gaussian in hydrodynamic
simulations of transonic compressible turbulence.  The centroid
velocity pdf for a 3D-forced MHD simulation given by Padoan et al.
(1999) looks Gaussian at velocities above the mean but has a fat
tail at small velocities. A detailed simulation study of the
centroid-velocity pdf was presented by Klessen (2000), who
examined driven and decaying hydrodynamic simulations with and
without self-gravity at various Mach numbers. The centroid pdfs
are nearly Gaussian, but the 3D pdfs were not discussed. The only
3D (not centroid) velocity pdf for an ISM-like simulation we know
of, a decaying MHD simulation with initial Mach number of 5
displayed by Ossenkopf \& Mac Low (2002, Figure 11a), is
distinctly exponential deep into the pdf core at their highest
resolution run; nevertheless they remark in the text that all the
pdfs are Gaussian.

Chappell \& Scalo (2001) found nearly-exponential tails in 2D
simulations of wind-driven pressureless (Burgers) turbulence, and
showed that the tail excesses persisted even in the absence of
forcing. They proposed that these excesses could be understood in
terms of the extreme inelasticity of the shell interactions in the
Burgers model, and speculated that the result could be more
general for systems in which interactions are inelastic, so that
kinetic energy is not a globally conserved quantity.  For example,
a Gibbs ensemble for particles that conserve mass and momentum, but
not energy, gives an exponential velocity distribution. Ricotti \&
Ferrara (2002) studied simulated systems of inelastically
colliding clouds driven by supernovae and showed how the velocity
pdf of the clouds approaches an exponential as the assumed
inelasticity becomes large.  Hyper-Gaussian velocity distributions
with exponential and even algebraic tails have also been observed
in laboratory and simulated inelastic granular fluids (Barrat \& Trizac 2002, Ben-Naim \& Krapivsky 2002, Radjai \&
Roux 2002). All
of these systems dissipate energy on all scales.

The reason for the fat tails may be the positive velocity
fluctuation correlations introduced by the inelasticity. This can
be seen by noting that successive velocities are positively
correlated for inelastic point particles conserving mass and
momentum but not energy. Then the velocity changes are not
independent and the central limit theorem does not apply (see
Mouri et al. 2003). Such correlations, resulting entirely from the
large range of dissipation scales, would be a fundamental
difference between incompressible and supersonic turbulence. If
true, then hyper-Gaussian pdfs need not be a signature of
intermittency (Klessen 2000).

\subsection{Turbulent ``Pressure''}
\label{sect:pres}

The idea of turbulent ``pressure'' is difficult to avoid because
of its convenience.  It has been used to generalize the
gravitational instability (e.g., Bonazzola et al. 1987,
V\'azquez-Semadeni \& Gazol 1995), to confine molecular clouds
(e.g., McKee \& Zweibel 1992, Zweibel \& McKee 1995), and to
approximate an equation of state (e.g., V\'azquez-Semadeni, Canto
\& Lizano 1998). Rigorous analysis (Bonazzola et al. 1992) shows
that turbulence can be represented as a pressure only if the
dominant scale is much smaller than the size of the region under
consideration (``microturbulence'').  In fact this is not the
case. The turbulent energy spectrum places most of the energy on
the largest scale, and because the spectrum is continuous, the
scale separation required for the definition of pressure
(Bonazzola et al. 1992) does not exist. Ballesteros-Paredes,
V\'azquez-Semadeni \& Scalo (1999) evaluated the volume and
surface terms in the virial theorem and showed that the kinetic
energy surface term is so large that the main effect of turbulence
is to distort, form, and dissolve clouds rather than maintain them
as quasi-permanent entities. Supersonic turbulence is also likely
to be dominated by highly intermittent shocks whose effect is
difficult to model as a pressure even with an ensemble average.

Simulations of turbulent fragmentation in self-gravitating clouds
show that turbulence suppresses global collapse while local
collapse occurs in cores produced by the turbulence (Sections
\ref{sect:gravity} and \ref{sect:clusters}).  This process does not
resemble pressure, however. Global collapse is avoided because the
gravitational and turbulent energies are transferred to separate
pieces that have smaller and smaller scales.

\subsection{Below the Collision Mean Free Path} \label{sect:smalltheory}

Scintillation observations suggest interstellar electron density
irregularities extend at a weak level down to tens or hundreds of
kilometers {\it Interstellar Turbulence II}), which is slightly
larger than the ion gyroradius ($m_icv_{th}/eB\sim 10^8/B(\mu G)$
cm) in the warm ionized medium and much smaller than the
ion-neutral ($10^{15}/n$ cm) and Coulomb ($10^{14}/n_i$ cm) mean
free paths at unit density, $n$ and $n_i$. Collisionless plasma
processes involving magnetic irregularities on small scales are
probably involved.

Collisionless plasmas behave like a fluid on scales larger than
the ion cyclotron radius. The equation of motion includes the
coupling between charged particles and the magnetic field. There
are also equations for mass and magnetic flux continuity and for
the heat flux.  The parallel and perpendicular components of the
pressure tensor that account for their gyromotions were included
by Chew, Goldberger \& Low (1956). A general review is in Kulsrud
(1983). Dissipation on very small scales is by cyclotron
resonance, which is not considered in these equations. Dissipation
on scales larger than the cyclotron radius is by Landau damping
and other processes, such as Ohmic and ambipolar diffusion. Landau
damping arises from a resonance between particle thermal speeds
and wave speeds in various directions. Proper treatment of Landau
damping in ISM turbulence is essential for understanding fast mode
waves that may scatter cosmic rays ({\it Interstellar Turbulence
II}). Landau damping could also modify the energy cascade because
it covers a wide range of scales.

Landau damping was included in the MHD equations by Snyder,
Hammett \& Dorland (1997). Passot \& Sulem (2003a) added
dispersive effects that are important on scales close to the ion
inertial length, $v_A/\Omega_i$, for Alfv\'en speed
$v_A=B/\left(4\pi\rho\right)^{1/2}$ and cyclotron frequency
$\Omega_i=eB/mc$.  They used the equations of continuity and
motion plus Ohm's law for the time derivative of the perturbed
field ${\bf b}$:
\begin{equation}
{{\partial {\bf b}}\over{\partial t}}-\nabla\times\left({\bf
u}\times {\bf b}\right)= -{1\over{{\hat
\Omega}_i}}\nabla\times\left({1\over\rho}\left(\nabla\times {\bf
b}\right)\times {\bf b}\right),\end{equation} where the quantities
are normalized to the ambient field and Alfv\'en speed, and ${\hat
\Omega}_i=\Omega_i L /v_A$ is the normalized ion cyclotron
frequency for system scale $L$.  The last term is the Hall term,
which comes from the inertia of the ions as the magnetic field
follows the electrons.  Pressure and density can be related by an
equation for the heat flux (Passot \& Sulem 2003a) or an adiabatic
power law on scales larger than the mean free path (Laveder,
Passot \& Sulem 2001). The result is a system of equations for
Hall-MHD. An interesting instability appears in this regime
resulting in a collapse of gas and field into thin helical
filaments (Laveder, Passot \& Sulem 2001, 2002). Because the
instability still operates in the collisionless regime (Passot \&
Sulem 2003b), such filaments may account for some of the elongated
structure seen by ISM scintillation on scales smaller than the
collision mean free path.

\subsection{MHD Turbulence Theory: Power Spectra}
\label{sect:powerspectra}

The hydrodynamic turbulent cascades and intermittency effects
discussed in Sections \ref{sect:scale} and \ref{sect:interm} have
analogs in the magnetohydrodynamic case. An important difference
between MHD and hydrodynamic turbulence arises because the
magnetic field gives a preferred direction for forces. Turbulence
in the solar wind (Matthaeus, Bieber \& Zank 1995) and on the
scale of interstellar scintillations is anisotropic with larger
gradients of density perpendicular to the field. Shebalin et al.
(1983) showed from incompressible numerical simulations with a
background magnetic field that the energy spectrum of velocity
fluctuations parallel to the field is steeper than perpendicular,
which suggests a diminished cascade in the parallel direction and
increasing anisotropy on small scales.  The suppression of the
parallel cascade depends on the strength of the mean field,
increasing for stronger fields (M\"uller, Biskamp \& Grappin
2003). In the extreme case, there is no cascade at all in the
parallel direction, leading to purely 2D or quasi-2D turbulence
(Zank \& Matthaeus 1993, Chen \& Kraichnan 1997, Matthaeus et al.
1998). Parallel structures at high spatial frequencies passively
follow the longer wavelengths in incompressible MHD without
cascading to the dissipation range (Kinney \& McWilliams 1998).

Incompressible MHD turbulence can be characterized in terms of
interactions between three shear-Alfv\'en wave packets (Montgomery
\& Matthaeus 1995, Ng \& Bhattacharjee 1996, Galtier et al. 2000),
which satisfy the wave number and frequency sum conditions, ${\bf
k}_1+{\bf k}_2={\bf k}_3$ and $\omega_1+\omega_2=\omega_3$, to
conserve momentum and energy. When the interacting waves are
Alfv\'en waves, $\omega=k_\|v_A$, and when they are oppositely
directed, ${\bf k}_{1}$ and ${\bf k}_{2}$ have opposite signs.
Then solutions exist only when the parallel Fourier modes have
energy at $k_{1\|}=0$ or $k_{2\|}=0$, which correspond to
long-wavelength field-line wandering. Also, when $k_\|=0$, there
is no cascade in the parallel direction (e.g., Bhattacharjee \& Ng
2001, Galtier et al. 2002). In this limit of incompressible or
weakly compressible MHD with a strong mean field, the transverse
velocity scales with wave number as $v_\bot\propto k^{-1/2}$ and
the 1D transverse energy scales as $E_\bot(k)\propto k^{-2}$ (Ng
\& Bhattacharjee 1997, Goldreich \& Sridhar 1997, Galtier et al.
2000).

Velocity and energy scalings with wave number may be obtained
from heuristic arguments that give physical insight to turbulence
theory (Connaughton, Nazarenko \& Newell 2003). There are two
important rates, the eddy interaction rate, $\omega_{int}$, and
the cascade rate, $\omega_{cas}$. These rates are related by the
number $N$ of interactions the fluid has to experience before the
energy at a certain wave number cascades,
$\omega_{cas}=\omega_{int}/N$. For weak interactions, the velocity
change per interaction, $\delta v$, is small and the number
required is determined by a random walk: $N\sim\left(v/\delta
v\right)^2$. For strong interactions, each one is significant, so
$\delta v\sim v$ and $N\sim1$. The velocity change comes from the
integral over the equation of motion for an interaction time:
$\delta v=\left(dv/dt\right)\omega_{int}^{-1}$. The most important
acceleration is the inertial term, ${\bf v}\cdot\nabla{\bf v}$,
giving $dv/dt\sim kv^2$ for wave number $k$ and velocity $v$ in the
cascading direction.

When $\delta v$ has this form, the interaction may be viewed as
consisting of three waves; three-wave systems have a constant
flux of energy, which is the cascade over wave number (Zakharov,
L'vov \& Falkovich 1992). Some systems such as gravity waves in
deep water (Pushkarev, Resio, Zakharov 2003) have stronger four-wave
interactions, and these conserve both energy and wave-action.
Their velocity change is given by $\delta
v=\left(d^2v/dt^2\right)\omega_{int}^{-2}$.

The last step in the derivation of the energy spectrum for weak
turbulence is to assume a constant energy flux in wave number
space, where energy is the square of the perturbed velocity or the
summed squares of the velocity and the perturbed field:
$\epsilon=v^2\omega_{cas}$. Combining terms for three-wave
interactions, we get,
\begin{equation}
\epsilon={{v^4k^2}\over{\omega_{int}}}={\rm
constant.}\label{eq:eps}\end{equation} For weak nearly
incompressible turbulence the interaction consists of oppositely
directed ``waves'' or Fourier components moving at the Alfv\'en
speed. The interaction rate is $\omega_{int}=k_\|v_A$ for parallel
wave number $k_\|$ and Alfv\'en speed $v_A$.  When the field is
strong, $k_\|\sim$constant so $\omega_{int}\sim$constant, giving a
transverse velocity scaling $v_\bot\propto k_\bot^{-1/2}$ from
equation (\ref{eq:eps}).   The energy spectrum follows from the
relation $\int E(k)dk=v^2$, so that $E(k)\propto v^2/k\propto
k^{-2}$. This is the result obtained by Ng \& Bhattacharjee (1997)
and others.

The Kolmogorov spectrum follows
from an isotropic picture in which the wave number and velocity of
the incoming perturbations are the same as those leaving the
interaction as part of the cascade. Then $\omega_{int}=kv$ and
$v^3k=$constant, giving $E(k)\propto k^{-5/3}$.

Goldreich \& Sridhar (1997) proposed that weak MHD turbulence is
irrelevant in the ISM because it quickly strengthens in the
cascade.  They suggested that ISM turbulence is usually strong,
and in this case there is a simplification that can be made from a
critical balance condition, $\omega_{cas}=\omega_{int}$ (Goldreich
\& Sridhar 1995). This condition gives a parallel cascade and
Kolmogorov scaling for transverse motions because if
$\epsilon=v_\bot^4k_\bot^2/\omega_{cas}$ from equation
(\ref{eq:eps}) and $\omega_{cas}=k_\bot v_\bot$, then the energy
flux is $\epsilon=v_\bot^3k_\bot$. Goldreich \& Sridhar reasoned
that inequalities between $\omega_{cas}$ and $\omega_{int}$ would
lead to changes in the interaction process that would restore
equality. If $\omega_{cas}>\omega_{int}$, for example, then local
field line curvature would decrease in time as the transverse
energy leaked away without adequate replacement from the parallel
direction. The field lines would then be more easily bent by the
next incoming packet.  An important assumption for their model is
that the wave interactions that dominate the energy transfer are
local in wave number space (see Lithwick \& Goldreich 2003).

For {isotropic} turbulence with a fixed incoming velocity, such as
an Alfv\'en speed, $\omega_{int}=kv_A$. Then one of the $k$'s
cancels from the numerator in Equation \ref{eq:eps}, but a
velocity does not cancel, giving $v^4k=$constant and $E(k)\propto
k^{-3/2}$.  This is the scaling suggested by Iroshnikov (1964) and
Kraichnan (1965) for magnetic turbulence, before the anisotropy of
this turbulence was appreciated. Recent studies of 2D MHD
turbulence show Iroshnikov-Kraichnan scaling also (Politano,
Pouquet \& Carbone 1998; Biskamp \& Schwarz 2001; Lee et al.
2003). Politano et al. actually got $\zeta_4\sim1$, which
corresponds to $<v^4>\propto k^{-1}$, but they got
$\zeta_2\sim0.7$, which gives about the Kolmogorov 1D energy
spectrum, $k^{-1.7}$.  The nonlinear dependence of $\zeta_p$ on
$p$ is the result of intermittency.

Iroshnikov-Kraichnan scaling and a suppressed parallel cascade
occurs in incompressible MHD turbulence if the energy transfer
among modes is dominated by interactions between waves with very
different sizes (Nakayama 2002). It may also apply in 2D MHD
turbulence with such nonlocal interactions (Pouquet, Frisch \&
L\'eorat 1976). The issue of locality in wave number space is not
understood even in hydrodynamic turbulence (Yeung, Brasseur \&
Wang 1995) although local transfer is commonly assumed. A study of
the degree of cancellation of long-range three-wave interactions
in incompressible hydrodynamic turbulence by Zhou, Yeung \&
Brasseur (1996) shows that nonlocal interactions cause anisotropy
at small scales, and the effect may increase with the scale
separation. This suggests that long-range dynamics persist at
large Reynolds numbers, in which case current MHD simulations may
not have the resolution to capture the effect.

Kolmogorov $k^{-5/3}$ scaling changes over to Iroshnikov-Kraichnan
$k^{-3/2}$ scaling as the mean field gets stronger, all else being
equal. M\"uller, Biskamp \& Grappin (2003) found this from
simulations of incompressible turbulence using a spectral code of
size $512^3$. As the transition to $k^{-3/2}$ occurred, the
inertial range parallel to the mean field became shorter
indicating an inability to cascade in this direction, as mentioned
above. Dmitruk, G\'omez \& Matthaeus (2003) also found a spectral
slope that depends sensitively on conditions. They considered
incompressible turbulence driven at speed $v_d$ at two opposing
boundaries of an elongated box measuring $L_\bot\times
L_\bot\times L_\|$. When the ratio of the Alfv\'en propagation
time along the field, $L_\|/v_A$, to the stirring time
perpendicular to the field, $L_\bot/v_d$, was large, the
relatively rapid stirring produced more small-scale structure and
a shallow energy spectrum. When the ratio was small, the spectrum
was steep.

The energy spectrum of weak nonmagnetic turbulence driven in a
rapidly rotating medium also shows Iroshnikov-Kraichnan scaling in
a direction perpendicular to the spin axis (Galtier 2003).

Scintillation observations ({\it Interstellar Turbulence II})
suggest the energy spectrum of density fluctuations is close to
Kolmogorov. This limits the range of possible models in these
applications. Also, the relevance of scaling laws derived under
the assumption of weak turbulence ($N=\left(v/\delta
v\right)^{2}>>1$) is questionable. For these reasons, there is
continued interest in the critical balance model of strong
turbulence discussed by Goldreich and collaborators.  We review
their proposals and the related numerical simulations next.

\subsection{The Anisotropic Kolmogorov Model}
\label{sect:goldreich}

Goldreich \& Sridhar (1995) proposed that interstellar turbulence
on small scales results from nonlinear interactions between shear
Alfv\'en waves in an incompressible, ionized medium. The energy
spectrum was determined from a critical balance condition that the
wave interaction rate in the parallel direction, $\omega_{int}\sim
k_\|v_A$, is comparable to the cascade rate in the perpendicular
direction, $\omega_{cas}\sim k_\bot v_\bot$. This condition gives
a cascading energy flux $\epsilon\sim v_\bot^2\omega_{cas}\sim
v_\bot^3k_\bot$, which leads to a scaling relation for
perpendicular motions, $v_\bot/v_A\sim
\left(\lambda_\bot/L\right)^{1/3}$, and a cascade in the parallel
direction, making $k_\| L\sim\left(k_\bot L\right)^{2/3}$. Here
$L$ is the scale at which extrapolated turbulent motions would be
isotropic and comparable to the Alfv\'en speed, $v_A$.

The energy spectrum is related to the velocity scaling law as
$v^2= \int E(k)dk$.
Cho, Lazarian \& Vishniac (2002a)
found from 3D MHD simulations that the energy spectrum for
parallel motions is $E(k_\|) \sim k_\|^{-2}$. They also obtained
$k_\|\propto k_\bot^{2/3}$ as above and fit the 3D power spectrum
to
\begin{equation}P(k_\bot,k_\|)=\left(B_0^2/L^{1/3}\right)k_\bot^{-10/3}
\exp\left(-L^{1/3}k_\|/k_\bot^{2/3}\right)\end{equation} for
unperturbed field strength $B_0$.

In this model, perturbations get stronger as they cascade to
smaller scales, and they get more elongated with
$\lambda_\|/\lambda_\bot\sim\left(2\pi L/\lambda_\bot
\right)^{1/3}$ increasing for smaller $\lambda_\bot$
($\lambda_\|/\lambda_\bot\sim1000$ on the smallest scales). These
local relations also follow if the local anisotropy,
$k_\|/k_\bot$, is proportional to the ratio of the local perturbed
field to the total field (Matthaeus et al. 1998; Cho, Lazarian \&
Vishniac 2003a). The global relation between $\lambda_\|$ and
$\lambda_\bot$, averaged over a large scale, can actually be more
isotropic, $\lambda_\bot\sim \lambda_\|$, if the local field lines
bend significantly on the small scale (Cho, Lazarian \& Vishniac
2002a).

\begin{figure}
\plotone{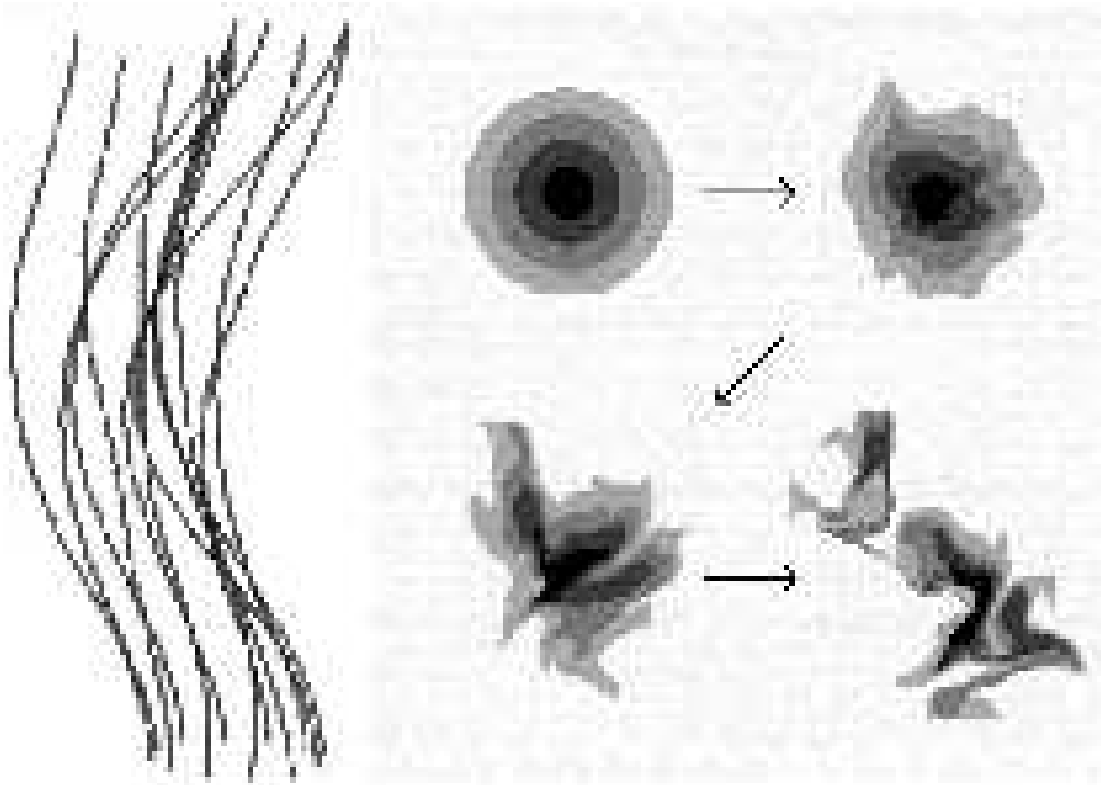} 
\caption{({\it left}) Distorted
field lines with a downward propagating wave, and ({\it right})
distortion of a bulls-eye pattern moving upward along these lines.
From Maron \& Goldreich (2001). {\it (for slightly higher resolution, see paperI-f4.gif)}}\label{fig:waves}
\end{figure}

A schematic diagram of a turbulent cascade is shown in figure
\ref{fig:waves}, from Maron \& Goldreich (2001). Wave packets
travel along the field and become distorted as the lines of force
interchange with other lines containing different waves. A density
pattern gets distorted by this motion too.

Electron density fluctuations in the theory of Goldreich \&
Sridhar (1995) are a combination of entropy fluctuations, i.e.,
temperature changes with approximate pressure equilibrium (Higdon
1986), and slow-mode wave compressions (Lithwick \& Goldreich
2001). Turbulence distorts and divides the large-scale density
structures into smaller structures, giving them the same power
spectrum as the velocity.  This is ``passive mixing'' if the
density irregularities are weak and have little back reaction on
the velocity field.

Passive mixing is a key component of the Goldreich et al. model
for which essentially all of the dynamics comes from
incompressible waves. Numerical simulations confirm the small
degree of coupling between these Alfv\'en shear modes and the
compressional modes, which are the fast and slow magnetosonic
modes (Section 4.1 in {\it Interstellar Turbulence II}). Cho \&
Lazarian (2002b) considered the low $\beta=P_{thermal}/P_{mag}$
case and separated the shear, fast, and slow modes in a
compressible MHD simulation. The relative energy in the
compressible part grew from zero to only 5\%--10\% after three
crossing times, which implies the slow mode is weakly coupled to
the shear mode. For purely solenoidal driving, the Alfv\'en and
slow modes had $k^{-5/3}$ energy scaling for velocity and field,
and the slow mode had $k^{-5/3}$ scaling for the density; the fast
modes had all three variables scale like $k^{-3/2}$.  Cho \&
Lazarian (2003) did similar calculations for the high $\beta$ case
and got the same results for mode coupling and energy spectra.
Most of the density structure was from slow modes at low $\beta$,
but at high $\beta$ it came from a nearly equal combination of
slow and fast modes. Cho \& Lazarian (2003) showed that even for
highly compressible supersonic turbulence, the Alfv\'en mode,
which barely couples to the compression, cascades like an
incompressible fluid with an energy spectrum $\propto k^{-5/3}$.

These scalings are consistent with the isotropies measured for the
three modes (Cho \& Lazarian 2002b). Alfv\'en modes and their
passive slow counterparts have velocities that depend on the
propagation direction relative to the field (at low $\beta$). This
makes them anisotropic with $k_\|\propto k_\bot^{2/3}$, as
discussed above.  Fast modes have equal velocities in all
directions at low $\beta$, and the resulting isotropy makes
$k_\|\propto k_\bot$, as proposed by Kraichnan (1965). Cho \&
Lazarian (2002b) further reasoned that slow modes are passive at
low $\beta$ because they move very slowly relative to the Alfv\'en
waves ($a\cos\theta<<v_A$ for thermal speed $a$),
so the Alfv\'en waves can distort them easily.

Cho, Lazarian \& Vishniac (2002b) proposed that magnetic
irregularities should persist without corresponding velocity
irregularities below the viscous damping length ($L_K$, see
Section \ref{sect:be}), down to the scale at which magnetic
diffusion becomes important. Damping in the equation of motion
comes from a term $\nu\nabla^2 v$, and magnetic diffusion comes
from a term like $\eta\nabla^2 B$; the ratio $\nu/\eta$ is the
Prandtl number. Cho et al. considered large Prandtl numbers and
found that the kinetic energy between the viscous and the
diffusion lengths cascaded more steeply than the magnetic energy:
$E_v(k_\bot)\propto k_\bot^{-4.5}$ and $E_B(k_\bot)\propto
k_\bot^{-1}$. The latter relation implies that the perturbed
field, $B_\bot$, is independent of scale (using $kE_b(k)\sim
b_\bot^2$), presumably because the cascade time into the
pure-magnetic regime is equal to the cascade time at its outer
scale, which is the viscous scale (Cho, Lazarian \& Vishniac
2003b). In their compressible MHD simulations, Cho, Lazarian \&
Vishniac (2003b) and Cho \& Lazarian (2003) found density
fluctuations below the viscous length with the same power spectrum
as the field, $E_\rho(k_\bot)\sim k_\bot^{-1}$. More recently,
Lazarian, Vishniac \& Cho (2004) suggested that velocity
fluctuations below the viscous length can be driven by magnetic
fluctuations. Schekochihin et al. (2002) found a folded field line
structure on small scales in incompressible MHD simulations at
high Prandtl number, with the most sharply curved fields being the
weakest. They also found comparable magnetic and kinetic energy
densities and explained this result in terms of a back-reaction of
the field on the motions.

Maron \& Goldreich (2001) and Cho, Lazarian \& Vishniac (2002a)
studied the velocity structure functions in strong incompressible
MHD turbulence. Recall that in a power-law approximation, these
functions can be written
\begin{equation}S_p\left(\delta r\right)=<[{\bf v}\left({\bf r}+
\delta {\bf r}\right)-{\bf v}\left({\bf r}\right)]^p>\sim\delta
r^{\zeta_p}\end{equation} for two points separated by distance
$\delta {\bf r}$. For Kolmogorov turbulence without intermittency,
$\zeta_p=p/3$; with intermittency, $\zeta_p(p)$ is nonlinear.
Politano \& Pouquet (1995) derived $\zeta_p(p)$ as
\begin{equation} \zeta_p={p\over g}\left(1-x\right)+C\left[ 1-
\left(1-{x\over C}\right)^{p/g}\right]\label{eq:pp}\end{equation}
where $g$ is the exponent in the velocity relation $v_\bot\propto
k_\bot^{-1/g}$, $x$ is the exponent in the cascade rate,
$\omega_{cas}\propto k_\bot^{x}$, and $C$ is the codimension of
the dissipation region: $C=2$ for lines and $C=1$ for sheets (the
codimension is equal to the number of spatial dimensions minus the
fractal dimension of the structure). For nonmagnetic Kolmogorov
turbulence with intermittency, $g=3$, $x=2/3$, and $C=2$, giving
the expression for $\zeta_p(p)$ originally found by She \& Leveque
(1994).

Cho, Lazarian \& Vishniac (2002a) determined that strong turbulent
motions perpendicular to the local field have the same
$\zeta_p(p)$ dependence as nonmagnetic turbulence. Relative to the
{global} magnetic field, the scaling of the perpendicular velocity
structure function was different (Cho, Lazarian \& Vishniac
2003b), following instead a form with $C=1$ (sheet-like) as
suggested by M\"uller \& Biskamp (2000). In both cases, the
magnetic structure function had an index $\zeta_p$ that was lower
than the velocity structure function, suggesting that $B_\bot$ is
more intermittent than $v_\bot$. The structure function of
velocity parallel to the mean local field had $\zeta_p$ larger
than for $v_\bot$ by a factor of 1.5, which is consistent with the
elongated geometry of the turbulence, for which
$\lambda_\|^{3/2}\propto\lambda_\bot$ (Cho, Lazarian \& Vishniac
2002a).

Haugen et al. (2003) did $1024^3$ simulations of forced,
weakly-compressible, nonhelical MHD turbulence and found a
codimension of $C\sim1.8$ (nearly line-like).  The total energy
spectrum integrated over 3D shells in $k$-space is $\sim k^{-5/3}$
at small $k$ and $\sim k^{-1.5}$ at intermediate $k$ just before
the dissipation range. This flattening to $k^{-1.5}$ was
attributed to a bottleneck in the cascade and not an
Iroshnikov-Kraichnan inertial spectrum. In the saturated state,
70\% of the energy was kinetic but 70\% of the dissipation
occurred by magnetic resistivity.  Such Ohmic dissipation can
increase the temperature in a turbulent medium by an order of
magnitude (Brandenburg et al. 1996).

Simulations of compressible MHD turbulence with zero mean
field (Cho, Lazarian \& Vishniac 2003b) had $\zeta_p$ for velocity
the same as for incompressible turbulence, giving $C=2$, and they
had $\zeta_p$ for the field in global coordinates satisfying the
above expression with $C=1$. Maps of the high-$k$ field structure
in planes perpendicular to the mean field clearly showed this
sheet-like field geometry for both compressible and incompressible
cases. There was a transition from somewhat uniform magnetic waves
on large scales to highly intermittent sheet-like regions on small
scales.

\section{SIMULATIONS OF INTERSTELLAR TURBULENCE}
\label{sect:sim}

\subsection{Introduction}
\label{sect:simintro}

Numerical simulations of the hydrodynamical or MHD equations
provide the only means by which we can actually ``observe''
interstellar turbulence in action, although what we see depends on
the physical processes included, whether and how external forcing
is applied, the dimensionality of the simulations, the imposed
initial and boundary conditions, the discretization technique, and
the spatial resolution. The first simulations of what can now be
seen as ISM turbulence go back over 20 years (Bania \& Lyon 1980),
but it is only within the past 5 to 10 years that the field has
matured enough to concentrate on the inherently turbulent aspects
of the problem.

The first simulations of nonmagnetic and nonself-gravitating
supersonic turbulence at relatively high resolution were by
Passot, Pouquet \& Woodward (1988).  Higher resolution studies of
such transonic turbulence are given by Porter et al. (1999) and
Porter, Pouquet \& Woodward (2002). Their models are dominated by
filaments in both vorticity (Figure \ref{fig:porter}) and
divergence of the velocity field, and the filaments cluster into
larger filaments and sheet-like structures in the compressible
part of the flow, forming sheets and spirals in the vorticity
field (Porter et al. 2002). Transonic nonself-gravitating models
like this may illustrate the morphology of diffuse clouds and
small molecular clouds when magnetic and gravitational forces are
not important. They are also able to isolate the effects of
moderate compressibility from other phenomena.

The first nonmagnetic simulations to include gravity were by
Leorat et al. (1990) on a $128^2$ grid, and the first
nongravitating compressible simulations to include magnetic fields
in an ISM context were by Yue et al. (1993) on a $20^2$ grid. Over
the next decade, these works were greatly expanded to include many
aspects of the ISM, such as magnetic and gravitational forces,
realistic heating and cooling, stellar energy injection and
galactic shear. The first efforts in this direction were by Chiang
\& Prendergast (1985) and Chiang \& Bregman (1988), who modeled
star-formation feedback in a 2D ISM, and by Passot,
V\'azquez-Semadeni \& Pouquet (1995), V\'azquez-Semadeni, Passot
\& Pouquet (1995), and V\'azquez-Semadeni, Passot \& Pouquet
(1996), who emphasized the importance of cooling and the nature of
turbulent forcing, and explored the relation between magnetism and
the star-formation rate. These were the first simulations to
include magnetic fields, self-gravity, heating and cooling, and
Coriolis forces, and were meant to represent the relatively
large-scale ($\sim1$ kpc) diffuse neutral ISM. By the late 1990s,
3D MHD simulations of isothermal molecular clouds without
self-gravity were performed by several groups (e.g., Mac Low et
al. 1998; Stone, Ostriker \& Gammie 1998;  Mac Low 1999; Padoan \&
Nordlund 1999), and detailed comparisons with observations were
made by Padoan, Jones \& Nordlund (1997) and Padoan et al. (1998,
1999).  Some problems with isothermal simulations are summarized
in Section \ref{sect:global}.

These models and others mentioned below are important steps toward
understanding the complex interactions that occur in the ISM, and
they provide useful statistical comparisons with observations.
However, like all turbulence simulations, they have a fundamental
limitation from the lack of dynamic range over the full ISM
Reynolds number, $Re$. The number of degrees of freedom for 3D
incompressible turbulence scales like $Re^{9/4}$, meaning that a
simulation at $Re=10^6$ would require a resolution of $10^5$ zones
per dimension, far beyond the reach of present-day approaches (but
see the lattice Boltzmann method of Chen et al. 2003).
Generalizing to compressible self-gravitating MHD with physically
consistent driving agents and an equation of state makes the
proposition even more implausible. One possibility for molecular
clouds with low ionization fraction is that the energy dissipation
scale is set by ambipolar diffusion and not viscosity (Klessen,
Heitsch \& Mac Low 2000). Then the effective Reynolds number might
be (Zweibel \& Brandenburg 1997)
\begin{equation}R_{AD}\approx\left(L/0.04\;{\rm pc}\right)
M_A\left(x/10^{-6}\right)\left(n/10^3\;{\rm cm}^{-3}\right)^{3/2}
\left(B/10\;\mu{\rm G}\right)^{-1}\end{equation} where $L$ is the
large scale of interest, $M_A$ the Alfv\'en Mach number, $x$
the ionization fraction, $n$ the particle density, and $B$
the magnetic field strength. In this case the available resolution
of numerical simulations would be adequate for dense molecular
clouds. However this argument neglects the possibility that
ambipolar diffusion acts on small scales to produce even finer
structure (Brandenburg \& Zweibel 1994; Tagger, Falgarone \&
Shukurov 1995; Balsara, Crutcher \& Pouquet 2001), and it only
applies in well-shielded molecular clouds, not in the general ISM.

An example of the importance of resolution for turbulence studies
is the test of whether the energy flux through the inertial range
is constant.  This is the basis for Kolmogorov's model of
incompressible turbulence, and it has only recently been verified
in $2048^3$ simulations on the Earth Simulator (Kaneda et al.
2003). Higher resolution ($4096^3$) simulations also find a
constant energy flux, but suggest there may be a significant small
departure from Kolmogorov's predicted energy spectrum (Kaneda et
al. 2003). Another example is the emergence of new phenomena at
very large $Re$, such as the transition in the power-law scaling
of the flatness factor, or kurtosis, of the velocity derivative
pdf observed for incompressible turbulence by Tabeling \& Willaime
(2002) at Taylor scale Reynolds numbers $Re_\lambda$ larger than
attainable by current simulations. For compressible turbulence in
the cold ISM, Adaptive Mesh Refinement simulations with effective
resolution as large as $10^3\times10^3\times10^4$ have not
converged yet either: a twofold increase in resolution yielded an
increase in maximum density and a decrease in minimum temperature
by factors greater than five  (de Avillez \& Breitschwerdt 2003).

Simulations at imperfect resolution can still reveal much of the
underlying physics as long as the range of scales resolved is
large enough that the intricate spatial fluid distortions
resulting from nonlinear interactions are expressed. Resolving the
inertial range for incompressible turbulence requires a resolution
equivalent to more than $\sim500$ zones in each direction (perhaps
significantly more; Pouquet, Rosenberg, \& Clyne 2003)---something
that is now commonly available in 3D and easily obtained for
smaller dimensions. With adequate resolution, simulations reveal
geometrical and topological properties, whereas theories using
statistical closure assumptions wash out the phase information.
Simulations also allow many more opportunities for comparison with
observations than do phenomenological or statistical models.

Here we give a brief summary of what seem to be the most important
accomplishments of ISM turbulence simulations and the questions
that remain unanswered.
Details of numerical techniques and their associated
problems are not discussed here.  For example, the use of periodic
boundary conditions in solving the Poisson equation introduces
artificial tidal forces that can be important away from the
density maxima (Gammie et al. 2003). The use of hyperviscosity,
sensitivity to assumed initial conditions, differences between
simulations using different numerical techniques, the violation of
the Jeans condition found by Bate \& Burkert (1997) and Truelove
et al. (1997) in some self-gravitating simulations, and the
assumption of isothermality in some simulations, are also not
addressed. Systematic effects resulting from varying the
parameters (especially the mean magnetic field) are also omitted
(see Ostriker et al. 2001, Lee et al. 2003). Readers are referred
to reviews of some of these problems and more detailed
discussions of statistical properties in V\'azquez-Semadeni et
al. (2000), Ballesteros-Paredes (2003), Mac Low (2003), Nordlund \& Padoan (2003), Ostriker (2003),
and Mac Low \& Klessen (2004).

\subsection{Scaling Relations}
\label{sect:larson}

Probably no result has generated more research on ISM turbulence
than Larson's (1981) finding that the density and velocity
dispersion of molecular clouds scale with the size of the the
region as power laws, suggestive of scaling relations that are
found in incompressible turbulence. Simulations have examined
these relations in detail. The density-size relation found by
Larson (1981), which implies a constant column density, may be a
selection effect because a wide range of densities is present for
each scale in the simulations, whereas observations tend to pick
the strongest emission regions and select for a restricted range
of column density (Kegel 1989; Scalo 1990; V\'azquez-Semadeni,
Ballesteros-Paredes \& Rodriguez 1997; Ballesteros-Paredes \& Mac
Low 2002). However, Ostriker et al. (2001) reproduced Larson's
squared dependence of the mass on size.

The linewidth-size relation is also problematic.  Some studies
show little correlation whereas others have a correlation
dominated by scatter; power-law fits also vary depending on tracer
and type of cloud (see references in Section \ref{sect:diag}).
Even if there is a power-law with an exponent 0.4--0.6, the
interpretation is ambiguous. It was attributed to a field of shock
discontinuities by V\'azquez-Semadeni et al. (1997), considering
that the associated velocity discontinuities should give an energy
spectrum $E(k)\propto k^{-2}$ (Saffman 1971) and so $v^2\propto
k^{-1}$, which gives $v\propto L^{1/2}$.  The problem with the
shock model is that energy spectra found for supersonic
simulations are not necessarily $k^{-2}$ (Boldyrev, Nordlund \&
Padoan 2002a; Wada et al. 2002; Vestuto et al. 2003). Also, the
Kolmogorov spectrum for incompressible turbulence, $E(k)\propto
k^{-1.67}$, is not much different. Further, the proposed shocks
themselves have not been observed directly in real clouds.
Nevertheless, the linewidth-size relations in Ballesteros-Paredes
\& Mac Low (2002) and Kim et al. (2001) were not just virialized
motions as these simulations did not include self-gravity; it is
not clear whether self-gravity is required to get observed
absolute scale. Kudoh \& Basu (2003) obtained the Larson relations
in numerical simulations of self-gravitating clouds with internal
energy from magnetic waves. Other simulation studies of this
relation are in Ostriker et al. (2001) and Ballesteros-Paredes \&
Mac Low (2002).

One problem with the linewidth-size observations is that it is
impossible to identify physically coherent condensations using
only radial velocities and sky coordinates of position.  This was
first pointed out by Burton (1971) and Adler \& Roberts (1992) in
connection with velocity crowding along galactic spiral arms, and
then recognized for interstellar turbulence by Lazarian \& Pogosyan (2000), Pichardo et al.
(2000), and Ostriker et al. (2001).
Another aspect of the problem is that the physically relevant
correlation may be between linewidth and density (Xie 1997, Padoan
et al. 2001b), which gives $v\propto n^{-0.4}$ to $n^{-0.3}$ for
Larson-like scaling. For supersonic turbulence, a relation like
this makes sense because the densest regions have been decelerated
behind shock fronts, although shocks propagating in a turbulent
medium can increase the velocity dispersion behind them (Rotman
1991, Andreopoulos et al. 2000), which gives the opposite effect.

Another scaling relation is between magnetic field strength and
density, which appears flat at densities $<100$ cm$^{-3}$ and then
has an upper limit that rises as $B\sim n^{0.5}$ above that
(Troland \& Heiles 1986, Crutcher 1999; Bourke et al. 2001).
Pre-simulation work concentrated on cloud contraction, but MHD
simulations account for the power index and scatter without
monotonic contraction. Examples include 2D models for the diffuse
ISM with self-gravity, cooling, and {\it H II} region expansion
(Passot et al. 1995), isothermal 3D models with and without
gravity forced with Fourier modes (Padoan \& Nordlund 1999,
Ostriker et al. 2001, Li et al. 2003), and 3D simulations with
supernovae (Kim et al. 2001). Turbulent ambipolar diffusion could
significantly affect the relation and its scatter (Heitsch et al.
2004). Passot \& V\'azquez-Semadeni (2003) present a detailed
analysis of the problem.  Other considerations are discussed in
Section \ref{sect:mag}.

\subsection{Decay of Supersonic MHD Turbulence}
\label{sect:decay}

A major discovery of simulations is that supersonic MHD turbulence
decays in roughly a crossing time (based on the rms turbulent
velocity) regardless of magnetic effects and the discreteness of
energy injection (Mac Low et al. 1998; Stone, Ostriker \& Gammie
1998; Mac Low 1999; Padoan \& Nordlund 1999; Avila-Reese \&
V\'azquez-Semadeni 2001; Ostriker et al. 2001). Decay is faster
with cooling (Kritsuk \& Norman 2002b, Pavlovski et al. 2002)
because the average Mach number is larger. Pavlovski et al. fit
the energy to $E(t)=E_0\left(1+t/t_1\right)^{\eta},$ for
$t_1=L_{inj}/v_{rms}$ equal to the initial ratio of the mean
energy injection scale to the rms velocity, and $\eta\sim-1.3$.
For isothermal compressible MHD turbulence that is either super-
or sub-Alfv\'enic, $\eta\sim-1$ (e.g., Mac Low et al. 1998). This
type of decay may be dominated by the pure Alfv\'enic (shear)
modes that have this time dependence if the compressional modes
are barely coupled to the shear (see Figure 1a in Cho \& Lazarian
2002b). Thus, damping is not just through the compressional modes
as originally thought. Alfv\'en wave propagation may actually
delay the energy decay (Cho \& Lazarian 2003). In addition,
compressible modes can propagate and diffuse freely along the
magnetic field lines.

Incompressible nonmagnetic turbulence is also believed to decay as
a power law with time, with an exponent between about $-1.2$ and
$-1.4$. Several approaches using various assumptions about the
preservation of the spectrum or the nature of the low wave number
part of the initial spectrum (e.g., Saffman's invariant,
Loitsianskii's invariant, closures whose results depend on the
initial spectrum) give results in this same range (see summaries
in Hinze 1975, Lesieur 1990, Frisch 1995, Biskamp 2003). Biskamp
\& M\"uller (1999, 2000) show that for incompressible isotropic
MHD turbulence the total and kinetic energies vary as $t^{-1}$
when there is no magnetic helicity but the total energy varies as
$t^{-1/2}$ with the same $t^{-1}$ kinetic energy decay when
helicity is present and nearly conserved. The physical reason for
the similar time dependencies in the incompressible, compressible,
and MHD cases is unknown.

A closely-related result, that clouds are transient entities not
confined by thermal or intercloud pressure, was found by
Ballesteros-Paredes, V\'azquez-Semadeni \& Scalo (1999) and
Elmegreen (1999). These results, along with observations of short
star-formation times (Ballesteros-Paredes, Hartmann \&
V\'azquez-Semadeni 1999; Elmegreen 2000; Hartmann et al. 2001),
eliminate the old problem of how clouds could be supported for
long times in the presence of supersonic turbulence (Zuckerman \&
Evans 1974).

The details of shock energy dissipation in 3D supersonic
turbulence have been studied by Smith, Mac Low \& Heitsch (2000) and
Smith, Mac Low \& Zuev (2000) for decaying and uniformly
(Fourier) forced motions; the latter paper includes models with
self-gravity. Besides the frequency distribution of shock
velocities whose form is explained with a phenomenological theory,
Smith et al. find that in the {decaying} runs the dissipation
rate peaks at small Mach numbers.  For example, at rms Mach number
of 5, the power loss peaks around Mach 1 to 2, while at rms Mach
50 the peak is from Mach 1 to 4, decreasing with time in both
cases.  Most of the dissipation is from a large number of weak
shocks.  In {uniformly forced} simulations, however, the
dissipation occurs in a small range of high Mach number shocks,
peaking at greater than the rms value. This result emphasizes the
crucial role of the driving process in controlling interstellar
turbulence and line profiles.  For example the smoothness of line
profiles (see Section \ref{sect:diag}) may imply that cool clouds
spend most of their time in the decaying mode, with only sporatic
and/or spatially discrete forcing. It is notable that in both
cases the magnetic field had little effect on the distribution of
shock strengths or their dissipation rate. As stated by Smith, Mac
Low \& Heitsch (2000), magnetic pressure is not damping the
shocks but helps maintain them by transferring energy from the
strong magnetic waves to the shock waves.

\subsection{The Density Probability Distribution}
\label{sect:lognormal}

V\'azquez-Semadeni (1994) suggested that the probability
distribution of densities (the density pdf) is lognormal as a
result of the central limit theorem applied to a multiplicative
hierarchical density field. Several groups subsequently found that
the pdf should exhibit an approximately lognormal distribution if
the gas is isothermal (Ostriker, Gammie \& Stone 1999; Klessen
2000; Ostriker et al. 2001; Li et al. 2003), with quasi-power-law
tails if nonisothermal (Passot \& V\'azquez-Semadeni 1998; Scalo
et al. 1998; Nordlund \& Padoan 1999).  Recent simulations of
dense cluster formation (Li, Klessen \& Mac Low 2003) confirm the
approximately lognormal behavior for isothermal turbulence and the
increasingly prominent high-density tail that appears when the
equation of state is softer.  Li et al. (2003) find density pdfs with
negative skewness from shocks that evacuate large regions, and
positive skewness from self-gravity that makes clumps. The
large-scale central disk galaxy simulations by Wada \& Norman
(2001) find a surprisingly robust and invariant lognormal form of
the density pdf (above the average density) when the density field
is generated by turbulence that includes a variety of physical
processes.  At densities below the peak the pdf found by Wada and Norman departs significantly from a lognormal.

This important theoretical result is difficult to verify from
observations of column densities alone. Gaustad \& VanBuren (1993)
sampled the three-dimensional density distribution for fairly low
densities, $\sim1$ cm$^{-3}$, using 1800 OB stars as probes of the
local dust density. Their pdf resembles a lognormal but it can
also be fit by a power law above the mean. Berkhuijsen (1999)
measured the volume filling factor as a function of density for
ionized and neutral gas and obtained a power law.

The probability distribution function for column density was
studied theoretically by Padoan et al. (2000), Ostriker et al.
(2001), and V\'azquez-Semadeni \& Garc\'ia (2001) using numerical
simulations of isothermal MHD turbulence.  For the isothermal case
the spatial density distribution is log-normal, so the column
density distribution is similar except for blending, which depends
on the ratio of cloud depth to autocorrelation length. For a large
ratio, the central limit theorem gives a Gaussian distribution of
column densities and for a very large ratio the column density
becomes nearly constant over the face of the cloud
(V\'azquez-Semadeni \& Garc\'ia 2001).

\subsection{Energy Cascades}

One of the oldest ideas in turbulence theory is that energy
cascades in wave number space. Molecular cloud simulations often
inject energy at large scales, and this energy cascades to smaller
scales where dissipation occurs. In analytical work, the transfer
is usually assumed to occur locally in wave number space, i.e.,
between similar-sized regions (compare Section
\ref{sect:powerspectra}). In fact, it is not well understood why
turbulence cascades in a given direction or how important nonlocal
energy transfer is (see Zhou, Yeung \& Brasseur 1996). Good
examples of nonlocal transfers are in shocks, where large-scale
flows abruptly convert kinetic energy into atomic random motions
at the front, and superbubbles, where localized energy gets
transferred in a single step to much larger scales. In 2D
incompressible turbulence, the energy cascades to large scales
regardless of where it is injected; only the mean squared
vorticity (enstrophy) cascades down to the viscous scale. Recent
work (Kurien, Taylor \& Matsumoto 2003) even indicates that the
inertial range of incompressible turbulence is a dual cascade of
energy and helicity that generates a $k^{-5/3}$ energy spectrum at
intermediate wave numbers but $k^{-4/3}$ at large wave numbers.

How should interstellar turbulence cascade?  Energy could be fed
in from rotation on the largest scales, self-gravity on
intermediate scales, and individual stars or clusters on
intermediate to small scales. The key concept behind a direct
cascade to smaller scales is absent, i.e., that an inertial range
exists in which advection alone distorts the fluid elements while
conserving kinetic energy. In addition, ISM turbulence is highly
compressible, so there is no quadratic invariant conserved by the
advection operator; energy can transfer directly between kinetic
and thermal modes.

Energy may cascade in either direction or in both simultaneously.
Most simulations that observe a cascade to small scales force the
turbulence over the whole computational domain and often use
purely solenoidal fluctuations, which favor the direct cascade.
Recently, Wada et al. (2002) showed that for 2D inner galaxy
models, the energy injected by self-gravity on $\sim10$ pc scales
cascades to larger regions; this is probably not an artifact of
the 2D geometry because the system is strongly compressible and
does not have the conservation properties that give incompressible
2D turbulence an inverse cascade. Vestuto, Ostriker \& Stone
(2003) found an inverse cascade in MHD simulations where the
forcing was not at the largest scales. Christensson, Hindmarsh \&
Brandenburg (2001) found inverse cascade in decaying 3D MHD
turbulence with helical fields.  Because the range of wave number
space over which forcing by {\it H II} regions, supernovae, and
superbubbles occurs is extremely broad and the energy input rate
is relatively uniform, the energy flow between scales in ISM
turbulence remains essentially unknown. Strong forcing over a wide
range of scales could even remove intermittency effects (Biferale,
Lanotte \& Toschi 2003).

\subsection{Compressible versus Solenoidal Motions}

Any turbulent velocity field can be decomposed into a solenoidal
(rotational) mode and a compressible (potential or irrotational)
mode and each mode can feed the other (Sasao 1973). It is of great
interest to know the relative energy in each. Intuitively it might
seem that the result should scale with Mach number because at zero
Mach number the flow is purely incompressible and at infinite Mach
number it should be mostly compressible.  Some discussion of the
flow of energy between the two modes, based on the coupling of the
evolution equations for the vorticity and dilatation, is given by
V\'azquez-Semadeni et al. (1996) and Kornreich \& Scalo (2000).
Bataille \& Zhou (1999) and Bertoglio, Bataille \& Marion (2001)
used a closure method with simulations to predict that the
compressible to solenoidal energy ratio should increase with the
square of the rms Mach number for very small Mach numbers.
However, Shivamoggi (1997) used a statistical mechanical approach
and found that the advection operator should evolve toward an
equipartition between compressible and vortical modes.

What do simulations have to say about this ratio?  A major problem
with simulations, especially at the molecular cloud level, is the
use of large-scale forcing in Fourier space. This has the effect
of stirring the gas in the entire computational domain
simultaneously.  The situation in the ISM, at least at
intermediate and small scales, is likely to be different.
Artificial forcing is usually solenoidal also, whereas stellar
energy sources will mostly generate compressible fluctuations.
Generally, in the magnetic case, the solenoidal component is
maintained at finite levels by magnetic torsions but in the
nonmagnetic case it can become very small (V\'azquez-Semadeni,
Passot \& Pouquet 1996).

Nordlund \& Padoan (2003) explain their solenoidal/compressional
ratio of $2/1$ geometrically: Interacting flows generate shear in
two dimensions but compression is only normal to the intersection
plane. Other simulations have solenoidal/compressional ratios
ranging from less than unity (V\'azquez-Semadeni et al. 1996) to
$5-10$, depending on Mach number (Boldyrev, Nordlund \& Padoan
2002b, Porter et al. 2002---the quantity quoted in the text of
Porter et al. 2002 is $E_{Com}/E_{tot}$ whereas the quantity
plotted in their Figure 2 is $E_{Com}/E_{Vor}$). Most results are
controlled by the degree to which the forcing is solenoidal or
compressible, the initial value of their ratio (Porter et al.
2002), the rms Mach number, the magnetic field strength, and the
importance of cooling (V\'azquez-Semadeni et al. 1996). Vestuto,
Ostriker \& Stone (2003) found that the solenoidal/compressional
ratio increases with higher magnetic field strength, verifying an
effect present in simulations by Passot et al. (1995) and
V\'azquez-Semadeni et al. (1996).  The unforced simulations of
Kritsuk \& Norman (2003) have an initially high fraction of
turbulent energy in the compressible form following thermal
instabilities and pancake formation, and then the energy converts
into mostly solenoidal form as the turbulence decays owing to
baroclinic vorticity generation from shell instabilities. The
value of this ratio was unfortunately not given in papers whose
simulations included supernova energy input; those should be the
most realistic for the ISM.

\subsection{Filamentary Structure}

The prevalence of filaments in the ISM has never been adequately
explained by turbulence theory, although they are obviously
present in the majority of simulations. Some ISM filaments are not
from turbulence but are the edges of expanding shells, cometary
tails of shocked clouds, or shocked regions in spiral density
waves. However, filamentary structure appears elsewhere too, even
in regions without obvious pressure sources (Kulkarni \& Heiles
1988, Jackson et al. 2003), and it occurs with similar morphologies in
turbulence simulations. Prominent suggestions for the origin of
these turbulence filaments include oblique shocks or shock
intersections, expanding gas along magnetic field lines,
confinement by helical magnetic fields, cooling instabilities in a
flattened large-scale structure, vorticity tubes from solenoidal
motions, and ambipolar diffusion.

Filamentary structure is often the result of shock interactions
(Heitsch et al. 2001).  Even in the simulations by Balsara,
Ward-Thompson \& Crutcher (2001), who emphasized the role of
magnetic field alignments, filaments were still found in the
converging flows. However, filamentary structures are also
prevalent in low Mach number, nonmagnetic simulations (Porter,
Pouquet \& Woodward 2002), where solenoidal forcing keeps the
compressible energy low and the shocks rare; they dominate the
vorticity field of incompressible turbulence (Figure
\ref{fig:porter}).

\subsection{Thermal Instability and Thermal Phases}
\label{sect:thermal}

The idea that thermal instability drives ISM condensation to
distinct phases (Field, Goldsmith \& Habing 1969) has lost much of
its appeal as a result of simulations that explicitly allow the
instability to operate. V\'azquez-Semadeni, Gazol \& Scalo (2000)
and S\'anchez-Salcedo, V\'azquez-Semadeni \& Gazol (2002) showed
how turbulence smears the bimodal density distribution expected in
an ISM whose structure forms solely by thermal instability. The
nonlinear development of a thermal instability in an initially
quiescent medium generates turbulence that reduces the initial
signature of the instability (Kritsuk \& Norman 2002b).  These
results do not arise because of strong departures from thermal
equilibrium (which is usually a good approximation away from
shocks), but simply because there is so much more kinetic than
thermal energy in supersonic turbulence.

Thermal pressure equilibrium of clouds is of minor importance in a
turbulent ISM (Ballesteros-Paredes, V\'azquez-Semadeni \& Scalo
1999). A high fraction of the gas mass can be in the thermally
unstable regime.  In the instability model, the underheated
regions in the unstable regime are slightly denser and at lower
pressure than the overheated regions so they contract at the sound
speed to become denser and more underheated. In a turbulent
medium, however, the underheated regions are not necessarily at
low pressure because they are pushed around by the surrounding
flow (Gazol et al. 2001). Large unstable fractions have also been
found in high resolution simulations of supernova-driven
turbulence (de Avillez \& Breitschwerdt 2004).

These results explain the observed large fraction of interstellar
gas in the supposedly unstable regime (Dickey, Salpeter \& Terzian
1977; Heiles 2001; Kanekar et al. 2003) and they explain the much
broader range of observed pressures than expected in the thermal
instability model Jenkins 2004), as shown by Kim, Balsara \& Mac
Low (2001). Neither of these features can be explained by
nonturbulent thermal instability models with pressure equilibrium.

\subsection{Supernova-Driven Turbulence}

Numerical simulations that have sufficient dynamic range to
include supernovae (Gazol-Pati\~no \& Passot 1999; Korpi et al.
1999a,b; de Avillez 2000;   de Avillez \& Berry 2001; Kim, Balsara
\& Mac Low 2001; Wada \& Norman 2001; Shukurov et al. 2004)
produce a continuous distribution of physical quantities, unlike
the previously prevailing paradigm of discrete ISM phases (Section
\ref{sect:thermal}). Temperatures weighted by volume segregate
into quasidiscrete ranges because the time spent in each
temperature range is proportional to the inverse of the derivative
of the cooling function (Gerola et al. 1974).  The density
distribution is not bi- or multimodal, however, as in a true
multiphase ISM. Thus, filling fractions of gas with various
temperatures may be approximately stable over long times in a
statistical sense, but these should not be interpreted as phases
since no phase transition is involved. Shukurov et al. (2004) and
Sarson et al. (2003) find that the filling factor of hot gas is
dependent on the presence, strength, and topology of the magnetic
field since a strongly ordered field confines the expanding gas
produced by supernovae. In the models by Wada \& Norman (1999),
the hot gas is heated primarily by turbulence itself.

Comparisons between these studies are difficult because some
include self-gravity (Wada \& Norman 2001) and magnetic fields
(Gazol-Pati\~no \& Passot 1999) but are 2D, whereas others are 3D
and either include the magnetic field with no self-gravity (Korpi
et al. 1999a, Kim et al. 2001, Shukurov et al. 2004) or include
neither field nor gravity (de Avillez 2000, de Avillez \& Berry
2001). There are also significant differences in the adopted
spatial distribution of the supernova explosions.  For example,
Gazol-Pati\~no \& Passot (1999), Wada \& Norman (2001), and
Shukurov et al. (2004) allow the supernovae to explode only at
sites where the turbulence produces local conditions conducive to
star formation, whereas Kim et al. (2001) explode supernovae
randomly. There are additional differences concerning how this
energy is distributed in time (e.g., whether a wind is allowed
before the explosion).  Many of these studies concentrate on the
resulting distribution of warm and hot gas, so it is difficult to
see from the published results how the phenomenology of the dense
gas is affected; in Shukurov et al. (2004), the cold gas is
explicitly omitted.

Nevertheless, the wealth of structure found in these simulations
is impressive. de Avillez (2000), de Avillez \& Berry (2001), and
de Avillez \& Breitschwerdt (2004) modeled the 3D vertical
structure in a galaxy disk at high resolution and found an amazing
array of structure over 400 Myr, including a thin cold disk
extruding vertical ``worms'' and a thick frothy disk of warm
neutral gas, thin sheets representing superbubble interactions
connected by tunnel-like structures, small clouds resulting from
the interaction and breakup of worms and sheets, networks of hot
gas, and ``chimneys'' rising high above the plane (see also Rosen,
Bregman \& Norman 1993).

The evolution of supernova remnants and superbubbles in a 3D MHD
turbulent medium is substantially different than in a uniform
external medium. Balsara, Benjamin \& Cox (2001) showed how the
induced time variations of synchrotron emission in young remnants
might serve as a diagnostic for the ambient medium, and they
pointed out how shock amplification of turbulent motions might be
important for cosmic rays.  Korpi et al. (1999b) showed that
superbubble breakout from galactic disks is easier in a turbulent
medium because of deformation of the expansion.

\subsection{The Role of Self-Gravity in the ISM}
\label{sect:gravity}

Elmegreen (1993) suggested that dense clouds or cloud cores could
form in colliding turbulent gas streams and that gravitational
instabilities in the shocked regions would lead to collapse and
star formation. The formation of self-gravitating condensations in
converging streams was suggested before this (Hunter et al. 1986),
but it was uncertain whether the compressed regions in a
{turbulent} flow would be large enough and live long enough to
collapse gravitationally before they dispersed.
Ballesteros-Paredes, V\'azquez-Semadeni \& Scalo (1999) confirmed
that density maxima in turbulent flows coincide with abrupt
velocity jumps, indicating shock formation. Other simulations
showed that dense regions generally form by oblique stream
collisions and intersecting shocks, and that the most strongly
self-gravitating of these regions have enough time to collapse
significantly (see Nordlund \& Padoan 2003).  Accretion and
core-core collisions often follow with more and more of the gas
entering this gravitating state. This process may terminate in a
real cloud when the resulting star formation becomes intense
enough to push the remaining gas away, or when the Mach number
drops below one (Section \ref{sect:clusters}).

The ability of nongravitating turbulence to fragment the ISM or
the interior of a GMC down to stellar masses and below leads to a
new concept for the role of gravity in forming interstellar
structure and stars (see Mac Low \& Klessen 2004 for a
comprehensive review). Clouds and clumps are no longer seen as the
result of gravitational instabilities in a region containing a
large number of Jeans masses, but simply the result of supersonic
turbulence, whatever the source.  On kpc scales, this turbulence
may be driven by interstellar gravity through swing-amplified
spiral arms, so gravity is important, but much of the cloudy
structure on smaller scales, including whole GMCs and their cores
and clumps, could result entirely from ``turbulent
fragmentation,'' a term that dates back to Kolesnik \&
Ogul$^\prime$Chanskii (1990). Cloud structure also results from
pressurized shell formation, spiral arm shocks, and other
processes, but even in these cases, the hierarchical clumpiness
inside the clouds probably arises from supersonic turbulence.

The role of gravity in determining gas structure is small when the
motions are faster than the virial speed. This is the case for the
diffuse ISM and the smaller {\it H I} and molecular clouds (Heyer
et al. 2001), and it applies to most of the small clumps inside
clouds (Loren 1989, Bertoldi \& McKee 1992, Falgarone et al.
1992).  Self-gravitating non-magnetic simulations by Klessen
(2001) confirm that the fraction of clumps with significant
self-gravity increases with mass. Many cloud properties can be
matched by MHD simulations without gravity, including the
statistical properties of extinction and intensity, the magnetic
field strength--density correlation, the size-linewidth relation,
and the absolute value and size independence of core rotation
(Padoan \& Nordlund 1999; Kim, Balsara \& Mac Low 2001; Ostriker
et al. 2001; Ballesteros-Paredes \& Mac Low 2002; Nordlund \&
Padoan 2003). The appearance of Bonner-Ebert density profiles does
not imply self-gravity either because turbulent condensations can
have this property even when they are not in equilibrium
(Ballesteros-Paredes, Klessen \& V\'azquez-Semadeni 2003).
Quasihydrostatic structures are actually difficult to produce by
turbulent fragmentation (Ballesteros-Paredes, V\'azquez-Semadeni
\& Scalo 1999). Moreover, the mass functions of dense cores
resemble the stellar initial mass function (IMF) in both
nongravitating (see Nordlund \& Padoan 2003) and gravitating (Li,
Klessen \& Mac Low 2003) simulations because the dense local
regions that can collapse were generated by turbulent
interactions, not gravitational instabilities. Self-gravity is
mainly important for the dense cores of turbulent regions, whereas
the large-scale structure is decoupled from these cores (Ossenkopf
et al. 2001).

If turbulence is viewed as preventing global collapse, then it is
not because of some ``turbulent pressure.'' In fact the conditions
required for turbulence to be represented as a pressure are quite
severe (Bonazzola et al. 1992), requiring a scale separation for
turbulent fluctuations (Section \ref{sect:pres}). Structures
generated by turbulence, even if bound, can be immune to global
collapse because the gravitational and turbulent energy is
transferred quickly to motions of substructures on smaller and
smaller scales, and the densest of these small structures does the
collapsing instead of the whole cloud. Collapse of
turbulence-compressed cores requires a low effective ratio of
specific heats $d\log P/d\log\rho=\gamma<2\left(1-1/n\right)$,
where $n=3,2,$ or 1 for spherical, planar, or filamentary
compressions, so the gravitational energy change exceeds the
compressional energy change (V\'azquez-Semadeni, Passot \& Pouquet
1996)

\subsection{Formation of Star Clusters and the IMF}
\label{sect:clusters}

\begin{figure}
\plotone{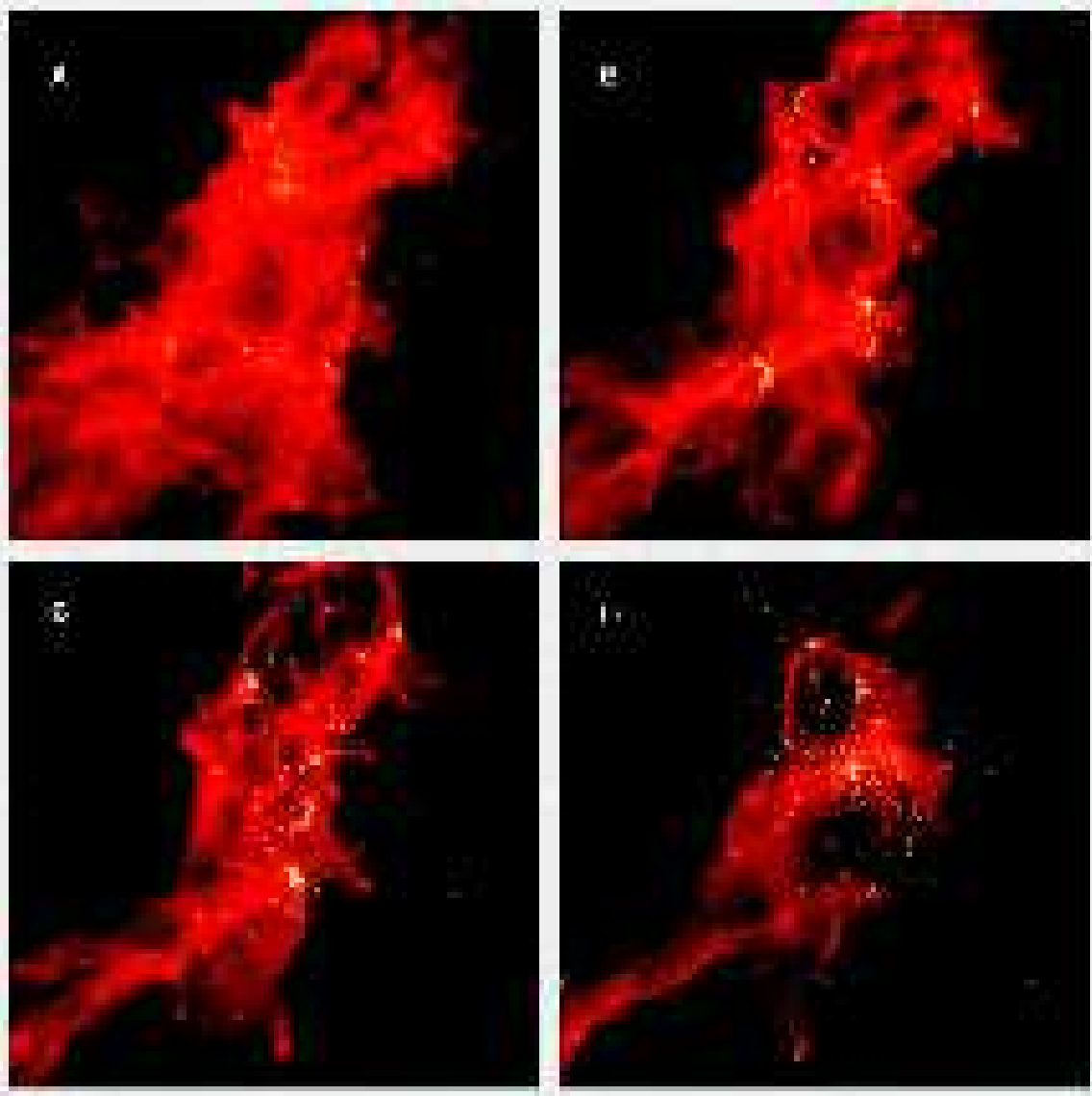} 
\caption{Four time steps in an
SPH hydrodynamic simulation with self-gravity and $5\times10^5$
fluid particles. Collapsing cores that represent stars are
replaced by sink particles in the simulation and are shown here by
white dots. Most stars form in a very dense gaseous core and then
get ejected into the general cluster field by two-body
interactions. Objects ranging in mass from brown dwarfs to
$\sim30$ M$_\odot$ stars are made (from Bonnell et al. 2003.) {\it (for slightly higher resolution, see paperI-f5.gif)}}
\label{fig:clu}\end{figure}

One of the holy grails of ISM simulations has been to follow the
collapse and fragmentation of a cloud to form a stellar cluster
and to calculate the resulting initial mass function of stars, a
goal that has been thwarted until recently by severe resolution
requirements. The first studies of this type did not include
self-gravity but showed that the resulting cores would collapse if
self-gravity were present, given their masses and densities (see
Nordlund \& Padoan 2003 and references therein). Self-gravitating
models illustrate the process in detail (Bonnell et al. 1997,
2003; Klessen, Burkert \& Bate 1998; Klessen \& Burkert 2000;
Klessen, Heitsch \& Mac Low 2000), including magnetic fields
(Heitsch, Mac Low \& Klessen 2001; Bate, Bonnell \& Bromm 2003; Li
et al. 2003). Figure \ref{fig:clu} shows a simulation result from
Bonnell et al. (2003).  The time sequence illustrates the collapse
of nonmagnetic gas into sink-particles (``stars''), competition
among these particles for gas, and the formation of disks,
binaries, and complex turbulent structures.

A major result of these studies is the demonstration that
turbulence can suppress global collapse whereas local collapse
still occurs in the cores that it forms. This result was found
even for magnetic simulations, where local collapse could be
suppressed only if the field was so strong that it supported the
whole cloud (Heitsch et al. 2001, Ostriker et al. 2001).

Klessen (2001) examined the mass spectrum of collapsing cores in
nonmagnetic, self-gravitating turbulence simulations and found
good agreement with the overall form of the stellar IMF (see
observations in Chabrier 2003).  It is interesting that roughly
similar clump mass spectra have been found in simulations
including self-gravity but neglecting magnetic fields (see also
Klessen \& Burkert 2001), simulations that include magnetic fields
but not self-gravity (Ballesteros-Paredes \& Mac Low 2002), and
simulations that include both (Gammie et al. 2003). Differences do
appear, but it is difficult to disentangle them from variations
resulting from numerical techniques and core definitions.

Gammie et al. (2003) presented a detailed discussion of the mass
spectra of gravitating and nonself-gravitating clumps in 3D
self-gravitating, decaying MHD simulations.  They found that the
slope of the high-mass part of the spectrum becomes shallower with
time, that the magnetic field strength does not affect the
spectrum significantly, and that the spectrum depends on how the
clumps are defined (Ballesteros-Paredes \& Mac Low 2002). The
dependence of the clump mass spectrum and the IMF on the equation
of state was studied by Li, Klessen \& Mac Low (2003). A
dependence of the substellar mass function on the power spectrum
of turbulence was found by Delgado-Donate, Clarke \& Bate (2004).

\subsection{Rotation and Binary Star Formation}

The rotational properties of molecular clouds and cores have been
investigated by several groups (e.g., Goldsmith \& Arquilla 1985,
Goodman et al. 1993, Caselli et al. 2002). Detectible gradients
range mostly from 0.3 to 3 km s$^{-2}$ pc$^{-1}$. Although the
average ratio of rotational to gravitational energy is small
($\sim0.03$), the nature of these gradients has been obscure.

Burkert \& Bodenheimer (2000) used synthetic velocity fields with
an assumed linewidth-size relation to show that even if turbulent
motions are random, the dominance of large-scale modes can lead to
velocity gradients that look like ordered rotation. The resulting
core rotational properties as functions of size and the spread in
the ratio of rotational to gravitational energies were in good
agreement with observations, including the spread in binary
periods, although the median angular momentum in the models was
about an order of magnitude larger than in the observations.
Gammie et al. (2003) studied properties of cores in 3D MHD
simulations of molecular clouds and also found angular momenta
larger than observed; they suggested higher resolution would
eventually allow simulations to probe turbulence-produced angular
momentum on the scale of binary star orbits. Fisher (2004) used a
semianalytic approach to show that the distribution of binary
periods was consistent with turbulence.

These results apply only to cores that are mildly subsonic.
Whether the small velocity gradients detected in some larger dark
clouds can be accounted for by turbulence alone remains to be seen
(but see Kim, Ostriker \& Stone 2003). Some gradients are probably
from pressurized motions. In any case, the angular momentum
problem that played a central role in most early discussions of
star formation (e.g., Mouschovias 1991) is beginning to disappear.
Angular momentum is still conserved in a turbulent ISM, but it
gets channeled mostly into vorticity structures as turbulence
distributes its energy among scales. It is unknown whether
turbulence can also explain the low levels of rotation in
prestellar cores (Jessop \& Ward-Thompson 2001).

Turbulence simulations are approaching the scales on which binary
star angular momenta are determined (Bate, Bonnell \& Bromm 2002;
Gammie et al. 2003). Even mild levels of turbulence during a
collapse can induce multiple star formation (Klein et al. 2003;
Goodwin, Whitworth \& Ward-Thompson 2004).

\subsection{Effects of Magnetic Fields on Interstellar
Turbulence} \label{sect:mag}

Most MHD simulations suggest that magnetic fields have a smaller
role in the ISM than previously expected. For example, they seem
unable to postpone the energy dissipation in a supersonically
turbulent medium (Section \ref{sect:decay}) and unable to prevent
local gravitational collapse unless they are strong (Heitsch et
al. 2001; Ostriker et al. 2001; however, see Kudoh \& Basu 2003).
Passot \& V\'azquez-Semadeni (2003) and Cho \& Lazarian (2003)
demonstrated that magnetic fluctuations do not act as an effective
pressure at high Mach numbers because of their lack of one-to-one
correspondence with density: The field acts more like a random
forcing of the turbulence. What magnetic fields appear to do is
reduce the core density contrast (Balsara, Crutcher \& Pouquet
2001) and slow down the local collapse rate (Heitsch et al. 2001).
This may not affect the overall star-formation rate much because
the timescale for core formation is much longer than the core
collapse itself. For turbulent fragmentation, star-formation rates
should be regulated on the largest scale where the dynamical time
is longest.  The marginal effects of magnetic fields on the clump
or core mass spectra and presumably the stellar IMF were discussed
in Section \ref{sect:clusters}.  Padoan \& Nordlund (1999)
suggested that magnetic fields are relatively weak in molecular
clouds, having energy densities less than turbulent so the motions
are super-Alfv\'enic (see review in Nordlund \& Padoan 2003).

Polarization observations show smooth magnetic fields, but this
does not mean turbulence is absent (see review by Ostriker 2003).
As discussed by V\'azquez-Semadeni et al. (1998), cloud formation
by turbulence can compress the field perpendicular to the
compression direction, making it elongated in the same direction
as the cloud; turbulent magnetic field-energy spectra are peaked
at small wave numbers in simulations, so the field should be
dominated by the largest scales, and the observed field
orientations are an average on the line of sight (Heiles et al.
1993, Goodman et al. 1990). Visual polarization also probes only a
small range of extinction.

Ambipolar diffusion is relatively unimportant in the overall
dynamics of molecular cloud turbulence at high Mach numbers
(Balsara, Crutcher \& Pouquet 2001). Indeed, observations now
suggest that ion-neutral drift is rapid at $\sim10^6$ cm$^{-3}$,
based on comparisons of HCO+ and HCN linewidths (Houde et al.
2002). The old idea that protostellar cores form on timescales
controlled by ambipolar diffusion may not apply.

Ion-neutral drift is not without consequences, however. The
ambipolar diffusion heating rate in molecular cloud turbulence
simulations (Padoan, Zweibel \& Nordlund 2000) can be orders of
magnitude larger on small scales than the background heating rate
by cosmic rays, and even the volume-averaged drift heating rate
can exceed the cosmic ray heating rate (Scalo 1977). Heating can
be important if dynamical effects are not because there is much
less thermal energy than kinetic energy in a supersonic flow.
Drift heating is also more important locally because it depends on
a high-order derivative of the magnetic field. Thus, turbulence
can enhance ambipolar diffusion by driving structures to smaller
scales (Zweibel 2002, Heitsch et al. 2004) and compressing the
field (Fatuzzo \& Adams 2002). Ambipolar diffusion also leads to
sharp filamentary structures (Brandenburg \& Zweibel 1994;  Mac
Low et al. 1995; Tagger, Falgarone \& Shukurov 1995), rather than
diffusive smoothing as once thought.

Turbulent cloud simulations (Gammie et al. 2003) also illustrate
why the observed clumps and cores in molecular clouds tend to be
prolate (Myers et al. 1991, Ryden 1996) with no preferred
orientation relative to the field (Goodman et al. 1990).
Quasistatic models with a slow contraction along field lines
predicted oblate and aligned cores.

Another interesting result concerning magnetic fields is the
possibility that they could control the accretion rate onto
protostellar cores or larger structures. Balsara, Ward-Thompson \&
Crutcher (2001) found with $256^3$ self-gravitating MHD
simulations that molecular cloud cores tend to accrete matter
nonuniformly along interconnecting magnetic filaments. They cite
evidence for this process in the S106 molecular cloud using
$^{13}$CO channel maps, submillimeter maps, and polarimetry.
Falgarone, Pety \& Phillips (2001) found six dense filaments
pointing toward the starless core L1512 extending out to about 1
pc, with evidence for the filamentary matter moving toward the
core. In this case the final masses of prestellar cores could be
controlled in part by the topology of the magnetic field.

Another magnetic effect is angular momentum braking, which has
been discussed for decades using idealized models of clouds and
fields. In a turbulent medium the magnetic field structure and its
connection to density condensations is complex, and there is no
guarantee that braking will occur.  V\'azquez-Semadeni, Passot \&
Pouquet (1996) and Vestuto, Ostriker \& Stone (2003) show
relatively more power in rotational and shear energy than
compressional energy as the field strength increases, contrary to
naive models of magnetic braking.

That magnetic braking does occur in a turbulent fluid with complex
fields has been shown in simulations of the magnetorotational
instability on a galactic scale (Kim, Ostriker \& Stone 2003).
The degree of braking of their largest coherent condensations may
explain the low specific angular momenta inferred for molecular
clouds.  This is a much larger scale than star-forming cores, but
the basic effect of angular momentum transfer along field lines is
analogous to the cases considered in idealized collapse models by
Mouschovias \& Paleologou (1979) and others.

A different approach was introduced by Cho and coworkers (see Cho
\& Lazarian 2002a,b; Cho, Lazarian \& Vishniac 2002a,b), Maron \&
Goldreich (2001), and M\"uller \& Biskamp (2003) using a
combination of theoretical calculations and simulations to address
the properties of anisotropic Kolmogorov turbulence. They
emphasized regimes that are more appropriate to the diffuse
ionized ISM than dense cold regions. Several interesting and
fundamental results have come from this, as reviewed in Section
\ref{sect:goldreich} and {\it Interstellar Turbulence II}.

\subsection{Turbulent Rotating Galaxy Disk Simulations}

Simulation codes are finally becoming sophisticated enough to
model at least the central regions of differentially rotating
galaxies, with simulations of 2 kpc x 2 kpc up to $4096^2$ (Wada
\& Norman 2001, Wada et al. 2002).  Various models include
self-gravity, cooling and background heating, and dynamical
heating from cluster winds and supernovae coupled to star
formation. The resulting density and temperature structures span
seven and five orders of magnitude, respectively.  The Wada et al.
(2002) models are not yet 3D, and they neglect magnetic fields,
but a lot of realistic structure is present anyway.  For example,
filaments form in oblique converging flows and strong local shear
(Wada et al. 2002), statistically stable temperature regimes
appear on large scales, the supernova rate fluctuates in recurrent
bursts from propagating-star formation, and the density pdf is
lognormal over four orders of magnitude above the mean. The models
also find that the energy spectrum and direction of the energy
flow in wave number space depends on the presence of star
formation feedback, the wavelength dependence of the disk
gravitational instability, and the assumed rotation curve for the
galaxy (compare Wada \& Norman 2001, Figure 18, and Wada et al.
2002, Figures 4 and 5). Wada, Spaans \& Kim (2000) also suggested
that turbulence alone could account for the kpc-sized holes seen
in other galaxies, independent of star formation.

\section{Summary and Reflections}

Interstellar turbulence has been studied using power spectra and
structure functions of the distributions of radial velocity,
emission, and absorption, using statistical properties of line
profiles, unsharp masks, and wavelet transforms, one-point
probability distribution functions of column densities and
velocity centroids, fractal dimensions and multifractal spectra,
and various other techniques including the Spectral Correlation
Function and Principal Component Analysis, which are applied to
spectral line data cubes.

The results are often ambiguous and difficult to interpret. The
density of the neutral medium seems to possess spatial
correlations over a wide range of scales, possibly from the
sub-parsec limit of resolution to hundreds of parsecs. Such
correlations probably reflect the hierarchical nature of
turbulence in a medium with a very large Reynolds number. The
power spectrum of the associated intensity fluctuations is a
robust power law with a slope between -1.8 and -2.3 for the Milky
Way, LMC, and SMC. A steeper slope has been obtained for Galactic
CO using Principle Component Analysis. However, no clear power
spectrum or autocorrelation function has been found yet for
centroid velocities as a function of spatial lag inside individual
clouds, even though they might be expected if the clouds are
turbulent.

Statistical properties of the ISM are difficult to recover with
only line-of-sight motions in projected cloud maps that have
limited spatial extent and substantial noise. Linewidths often
seem correlated with region size in an average sense, but there
are large variations between different regions and surveys.  At
the moment, this relation seems to be dominated by scatter.
Because the linewidth-size correlation is analogous to a second
order structure function, which is well-defined for incompressible
turbulence, the regional variations are difficult to understand in
the context of conventional turbulence theory. Certainly a case
could be made that ISM turbulence is not conventional: it is not
statistically homogeneous, stationary, or isotropic.

Considering the difficulty of observing three-dimensional
structure and motion in space, the value of statistical
descriptors lies primarily in their ability to make detailed
comparisons between observations and simulations. Many such
comparisons have been made, but a comprehensive simulation of a
particular region including many of the descriptors listed above
is not yet available.

Among the many differences between interstellar turbulence and
classical incompressible turbulence is the broad spatial scale for
energy input in space. In spite of the wide ranging spatial
correlations in emission and absorption maps, there is no analogy
with classical turbulence in which energy is injected on the
largest scale and then cascades in a self-similar fashion to the
very small scale of dissipation. In the ISM, energy is injected
over a wide range of scales, from kiloparsec disturbances in
spiral density waves, shear instabilities, and superbubbles, to
parsec-sized explosions in supernovae, massive-star winds and {\it
H II} regions, to sub-parsec motions in low-mass stellar winds and
stellar gravitational wakes, to AU-sized motions powered by cosmic
ray streaming instabilities. Estimates of power input indicate
that stellar explosions are the largest contributor numerically,
but this does not mean that other sources are unimportant on other
scales.

Energy is also dissipated over a wide range of scales.  Shock
fronts, ambipolar and Ohmic diffusion, Landau wave damping, and
viscosity in vortex tubes should all play a role. The geometrical
structures of the dissipating regions are not well constrained by
observations.  Energy is also dissipated in regions with little
density sub-structure, as in the ionized gas that produces
scintillation or along perturbed magnetic field lines that scatter
some cosmic rays. Viscosity and Landau damping can produce heat
from the tiny wave-like motions that occur in these regions (see
Interstellar Turbulence, Part II), although the nature of this
dissipation below the collision mean free path is not understood
yet.

Self-gravity makes interstellar turbulence more difficult to
understand than terrestrial turbulence. The contribution to ISM
motions from self-gravity appears to increase with cloud mass
until it dominates above $10^4$--$10^5$ M$_\odot$. Many selection
effects may contribute to this correlation, however. Self-gravity
is also important in the smaller clumps that form stars. What
happens between these scales remains a mystery.  We would like to
know how the distribution function of the ratio of virial mass to
luminous mass for ISM clouds and clumps varies with the spatial
scales of these structures. What fraction of clouds or clumps are
self-gravitating for each mass range? Does the apparent trend
toward diminished self-gravity on small scales turn around on the
scale of individual protostars?

The balance between solenoidal and compressible energy density in
the ISM varies with time. Compressibility transfers energy between
kinetic and thermal modes, short-circuiting the cascade of energy
in wavenumber space that occurs in a self-similar fashion for
incompressible turbulence. Consequently the ISM turbulent power
spectra for kinetic and magnetic energy are not known from
terrestrial analogies, and there is not even a theoretical or
heuristic justification for expecting self-similar or power-law
behavior. Numerical simulations make predictions of these
quantities, but these simulations are usually idealized and they
are always limited by resolution and other numerical effects.

A considerable effort has been put into modelling compressible MHD
turbulence under idealized conditions, i.e., without self-gravity
and without dispersed and realistic energy sources (e.g.,
explosions). Some models predict analytically and confirm
numerically a Kolmogorov energy spectrum transverse to the mean
field when the magnetic energy density is not much larger than the
kinetic energy density. These models also predict intermittency
properties identical to hydrodynamic turbulence in this
trans-field direction. Other models find steeper power spectra,
however. One has mostly solenoidal motions on large scales and
sheet-like dissipation regions on small scales. Another has more
realistic heating and cooling. At very strong fields, the
turbulence becomes more restricted to the two transverse
dimensions and then the energy spectrum seems to become flatter,
as in the Iroshnikov-Kraichnan model.  Overall the situation
regarding the energy spectrum of MHD ISM turbulence is unsettled.

Our understanding of the ISM has benefited greatly from numerical
simulations. In the 10 or so years since the first simulations of
ISM turbulence, numerical models have reproduced most of the
observed correlations and scaling relations.  They have
demonstrated that supersonic turbulence always decays quickly and
concluded that star formation is equally fast, forcing a link
between cold clouds and their energy-rich environments. They have
also predicted a somewhat universal probability function for
density in an isothermal gas, implying that only a small fraction
of the gas mass can ever be in a dense enough state to collapse
gravitationally -- thereby explaining the low efficiency of star
formation.

Simulations have demonstrated that magnetism does not support
clouds, prevent collapse, systematically align small clumps, or
act like an effective pressure. Magnetism can slow collapsing
cores and remove angular momentum on large and small scales, and
it may contribute to filamentary structures that control the
accretion rate of gas onto protostars. Simulations have also shown
that thermal instabilities do not lead to bimodal density
distributions as previously believed, although they probably
enhance the compressibility of the ISM and might contribute to the
turbulent motions.  The high abundance of atomic gas in what would
be the thermally unstable regime of static models is easily
explained in a turbulent medium.

Simulations have a long way to go before solving some of the most
important problems, however. Models are needed that include
realistic cooling, ionization balance, chemistry, radiative
transfer, ambipolar diffusion, magnetic reconnection, and
especially realistic forcing because it is not clear that any of
these effects are negligible. Detailed models should be able to
form star clusters in a realistic fashion, continuing the progress
that has already been made with restricted resolution and input
physics. Complete models need to include energy sources from young
stellar winds, ionization, and heating, and they should be able to
follow the orbital dynamics of binaries. Successful models should
eventually explain the stellar initial mass function, mass
segregation, the binary fraction and distribution of binary
orbital periods, the mean magnetic flux in a star, and the overall
star-formation efficiency at the time of cloud disruption.

Realistic simulations of galactic-scale processes are also
challenging because they should include a large range of scales
and fundamentally different physical processes in the radial and
vertical directions. Background galactic dynamics and the
possibility of energy and mass exchange with a three-dimensional
halo are also important.

The turbulent model of the ISM also raises substantial new
questions. Why is the power spectrum of ISM density structure a
power law when direct observation shows the ISM to be a collection
of shells, bubbles, comets, spiral wave shocks and other
pressurized structures spanning a wide range of scales? Do
directed pressures and turbulence combine to produce the observed
power spectrum or does one dominate to other? Does the input of
energy on one scale prevent the turbulent cascade of energy that
is put in on a larger scale?  In this respect, is ISM turbulence
more similar to turbulent combustion than incompressible
turbulence because of the way in which energy is injected?

Our current embrace of turbulence theory as an explanation for ISM
structures and motions may be partly based on an
overly-simplification of available models and a limitation of
observational techniques. This state of the field guarantees more
surprises in the coming decade.

{\it Acknowledgments:} We are grateful to J. Ballesteros-Paredes,
J. Dickey, N. Evans, E. Falgarone, A. Goodman, A. Lazarian, P.
Padoan, and E. V\'azquez-Semadeni for helpful comments on Section
2;  to C. Norman for a careful reading of Section 3; to A.
Brandenberg, A. Lazarian, P. Padoan, T. Passot, E.
V\'azquez-Semadeni, and E. Vishniac for helpful comments on
Section 4; and to A.G. Kritsuk, M-M. Mac Low, E. Ostriker, P.
Padoan, and E. V\'azquez-Semadeni for helpful comments on Section
5. We also thank J. Kormendy for helpful comments on the
manuscript.

\end{document}